\documentclass[sn-mathphys,Numbered]{sn-jnl}

\usepackage{graphicx}%
\usepackage{multirow}%
\usepackage{amsmath,amssymb,amsfonts}%
\usepackage{amsthm}%
\usepackage{mathrsfs}%
\usepackage[title]{appendix}%
\usepackage{xcolor}%
\usepackage{textcomp}%
\usepackage{manyfoot}%
\usepackage{booktabs}%
\usepackage{algorithm}%
\usepackage{algorithmicx}%
\usepackage{algpseudocode}%
\usepackage{listings}%
\usepackage{multirow}
\usepackage{subcaption}
\usepackage{graphicx}
\usepackage[normalem]{ulem}  
\def\rts{RT$_{60}$}

\begin{document}

\title[Article Title]{Exploring the Power of Pure Attention Mechanisms in Blind Room Parameter Estimation}


\author[1]{\fnm{Chunxi} \sur{Wang}}\email{chunxiwang@emails.bjut.edu.cn}

\author[1]{\fnm{Maoshen} \sur{Jia}}\email{jiamaoshen@bjut.edu.cn}

\author[1]{\fnm{Meiran} \sur{Li}}\email{lmran@emials.bjut.edu.cn}

\author[1]{\fnm{Changchun} \sur{Bao}}\email{baochch@bjut.edu.cn}

\author[2]{\fnm{Wenyu} \sur{Jin}}\email{wenyu.jin@ieee.org}

\affil[1]{\orgdiv{Beijing Key Laboratory of Computational Intelligence and Intelligent System
}, \orgname{Faculty of Information Technology}, \orgaddress{\street{Beijing University of Technology}, \city{Beijing}, \postcode{100124},  \country{China}}}

\affil[2]{\orgdiv{AcousticDSP Consulting LLC}, \orgaddress{\street{St Paul}, \state{MN}, \country{United States}}}



\abstract{Dynamic parameterization of acoustic environments has drawn widespread attention in the field of audio processing. Precise representation of local room acoustic characteristics is crucial when designing audio filters for various audio rendering applications. Key parameters in this context include reverberation time (\rts) and geometric room volume. In recent years, neural networks have been extensively applied in the task of blind room parameter estimation. However, there remains a question of whether pure attention mechanisms can achieve superior performance in this task. To address this issue, this study employs blind room parameter estimation based on monaural noisy speech signals. Various model architectures are investigated, including a proposed attention-based model. This model is a convolution-free Audio Spectrogram Transformer, utilizing patch splitting, attention mechanisms, and cross-modality transfer learning from a pretrained Vision Transformer. Experimental results suggest that the proposed attention mechanism-based model, relying purely on attention mechanisms without using convolution, exhibits significantly improved performance across various room parameter estimation tasks, especially with the help of dedicated pretraining and data augmentation schemes. Additionally, the model demonstrates more advantageous adaptability and robustness when handling variable-length audio inputs compared to existing methods.}




\maketitle

\section{Introduction}\label{sec1}

In recent years, there has been a growing focus on the dynamic parameterization of evolving acoustic environments. The parameters that describe local rooms or other acoustic spaces hold significance as they can be harnessed in the modelling and design of audio filters for a diverse range of applications. Understanding the specific acoustic properties of the surrounding room can be applied to improve speech signals and support dereverberation algorithms, ultimately improving word error rate for Automatic Speech Recognition (ASR) and the clarity of voice communication systems \cite{Zhang18,Wu17,Mohammadiha16}. Additionally, spatial sound reproduction systems could leverage this data to enhance their performance in tasks related to acoustic room equalization either using predefined filters \cite{Cecchi18,Jin15} or in an adaptive manner \cite{Jin16}. 

Furthermore, the successful realization of audio augmented reality (AAR) necessitates the seamless integration of virtual acoustic objects into the physical environment. This integration underscores the importance of achieving a harmonious alignment between the acoustical properties of virtual elements and the characteristics of the local space \cite{Neidhardt22}. In pursuit of this goal, a significant challenge lies in the accurate estimation of related acoustical parameters of a room to enhance the realism of immersive audio. Notably, Jot et al. \cite{Jot16} introduced the concept of a ``reverberation fingerprint", comprising the room's volume and its frequency-dependent diffuse reverberation decay time. This innovative concept was proposed to characterize rooms specifically for the realistic binaural rendering achievable with audio augmented reality headphones. It's worth noting that this fingerprint primarily focuses on the part of reverberation that is independent of the position, treating a room's acoustic characteristics in isolation from the orientation and directivity of sound sources and receivers.

Conventionally, room parameters like reverberation time (\rts) and direct-to-reverberant ratio (DRR) are typically obtained through a direct analysis of measured Room Impulse Responses (RIRs). Meanwhile, room volume is closely linked to the determination of a concept known as the ``critical distance." This critical distance is defined as the distance at which the direct and reverberant power components of a sound source become equal, effectively making the DRR reach 0 dB. In cases where we assume an ideal diffuse soundfield, the relationship between these parameters is mathematically described by Sabine's well-known equation \cite{Kuttruff16}:
\begin{equation}
    RT_{60}(b) \approx 0.16  \frac{V}{\alpha(b) \cdot S},
\end{equation}
where $S$ denotes the total area of the room’s surfaces and $\alpha(b)$ is the area-weighted mean absorption coefficient in octave band $b$.

In practical applications, conducting in-situ measurements of RIRs and determining the volumes of users' local acoustic spaces can often present significant challenges. A compelling alternative involves blind estimation of room acoustic parameters from audio recordings obtained using microphones, even when the sound sources are unknown and in the presence of background noise. The 2015 ACE challenge \cite{Eaton16} established a benchmark for blind estimation of \rts\ and DRR from noisy speech recordings. The leading systems in this challenge primarily relied on signal modeling-based approaches \cite{Prego15,Loellmann15}. Meanwhile, room volume estimation has long been formulated as a classification problem \cite{Moore14,Peters12}. Audio forensics systems described in \cite{Moore14,Peters12} make use of Mel-frequency cepstral coefficients (MFCC)-based features to identify the specific room associated with an environmental sound or speech recording, typically within a predefined closed set of possibilities.

Due to the recent advancements in Deep Neural Networks (DNNs), there is a growing trend to reframe the challenge of blind room acoustic parameter estimation as a regression problem. This approach leverages Convolutional Neural Network (CNN) models in combination with time-frequency representations, offering an increasingly relevant and effective solution. Gamper et al. \cite{Gamper18} introduced a CNN designed to directly estimate \rts\ from a four-second recording of reverberant speech. The experimental results demonstrate that this CNN outperforms other methods in the ACE challenge, offering both superior performance and higher computational efficiency. The same approach was also applied to blind volume estimation in \cite{Genovese19} and results show that it can estimate a broad range of volumes from real-measured data (with average estimated errors typically ranging from half to twice the actual values). CNN-based systems with similar methodologies have been put forward for the blind estimation of room acoustic parameters, utilizing either single-channel \cite{Bryan20,Gotz22,Saini23} or multi-channel speech signals \cite{Srivastava21}. These systems have showcased promising outcomes in terms of both accurate parameter estimation and resilience to temporal variations in dynamic acoustic environments. Notably, in contrast to the conventional approach of log-energy calculations for spectro-temporal features used in prior studies, Ick et al. \cite{Ick23} introduced a set of phase-related features. Their research demonstrated clear improvements in the context of estimating reverberation fingerprints for real-world rooms that had not been previously seen, highlighting the enhanced efficacy of this method.

CNNs are widely considered in the fore-mentioned approaches due to their suitability for learning two-dimensional time-frequency signal patterns for end-to-end modelling. CNNs can be extended by a recurrent layer to form convolutional recurrent neural networks (CRNN) that exploit sequential dependencies in the data \cite{Callens2020} and improve the capability of processing input sequences of variable length \cite{Deng2020}. To further enhance the capture of long-range global context, hybrid models combining Convolutional Neural Networks (CNNs) with self-attention mechanisms have yielded state-of-the-art results in a range of tasks, including acoustic event classification \cite{Kong20,Gong21} and various audio pattern recognition endeavors \cite{Li2018,Rybakov20}. Gong et al. \cite{Gong21ast} pushed the boundaries even further by introducing purely attention-based models for audio classification. Their creation, the Audio Spectrogram Transformer (AST), was evaluated on several audio classification benchmarks, achieving new state-of-the-art results. This underscores that CNNs may not always be essential in this particular context.


Building on the inspiration derived from the research presented in \cite{Gong21ast}, our study introduces a convolution-free, purely attention-based model for the blind estimation of acoustic room parameters by extending our previous work in \cite{Wang2023}. To the best of our knowledge, this marks the inaugural application of an attention-based system in the field of blind acoustic room parameter estimation. The proposed system utilizes Gammatone magnitude spectral coefficients as well as the low-frequency phase spectrogram as inputs and captures long-range global context, even in the lower layers of the model. Furthermore, to enhance system performance, we apply transfer learning through the use of a pretrained transformer model from ImageNet. For the evaluation of the proposed method, we curate a RIR corpus that includes publicly available RIRs, synthesized RIRs, and RIRs obtained through in-house measurements of real-world rooms. Experimental results clearly demonstrate the superiority of our proposed model when compared to CNN-based blind acoustic parameter estimation systems, particularly when dealing with previously unseen real-world rooms using single-channel recordings of variable length.

The remainder of the article is organized as follows.

 Section 2 introduces the construction of RIR datasets, including real-world and simulated datasets. Section 3 demonstrates the generation of audio data with reverberation and noise using constructed RIR datasets, followed by data preprocessing, augmentation, and feature extraction schemes for neural network training. Section 4 details the model structures of a CNN-based model, a CRNN-based model, the proposed attention-based systems. Section 5 conducts a comprehensive evaluation of the proposed system against state-of-the-art methods in various room parameter estimation tasks and its performance under variable-length inputs. Section 6 draws the conclusion.

 \section{Data generation pipeline}
Applying neural network methods to address blind room parameter estimation is a challenging task, as it generally requires a substantial amount of data. Since this task necessitates the need of having audio samples from rooms with various acoustic characteristics, manually creating a suitably diverse dataset would incur exorbitant costs and time. In this work, audio samples are created from public real-world RIR datasets, the BJUT Reverb dataset, and a room-simulation-based RIR dataset. 

\subsection{Public real-world RIR datasets}
In this study, six publicly available real-world RIR datasets that include 44 authentic rooms are considered, with the aim of encompassing a wide range of acoustic room parameters. 

The majority of the data targets at geometrically regular rooms, including spaces like offices, classrooms, and auditoriums/lecture halls. These datasets include the ACE Challenge dataset \cite{Eaton16}, the Aachen Impulse Response (AIR) dataset \cite{Jeub09}, the Brno University of Technology Reverb Database (BUT ReverbDB) \cite{Igor19}, the C4DM dataset \cite{Stewart10}, and the dEchorate dataset \cite{Carlo21}. Additionally, the OpenAIR dataset \cite{murphy2010openair:} primarily covers larger acoustic spaces, such as churches, nuclear reactor halls, and other substantial structures. As a result, a large variety of real-world room configurations with different volume parameters are incorporated.

Furthermore, \rts\ values vary widely, ranging from less than half a second to over ten seconds, and these values are calculated using the Schroeder method \cite{schroeder}. 

\begin{figure}
    \centering
    \includegraphics[height=4.69cm]{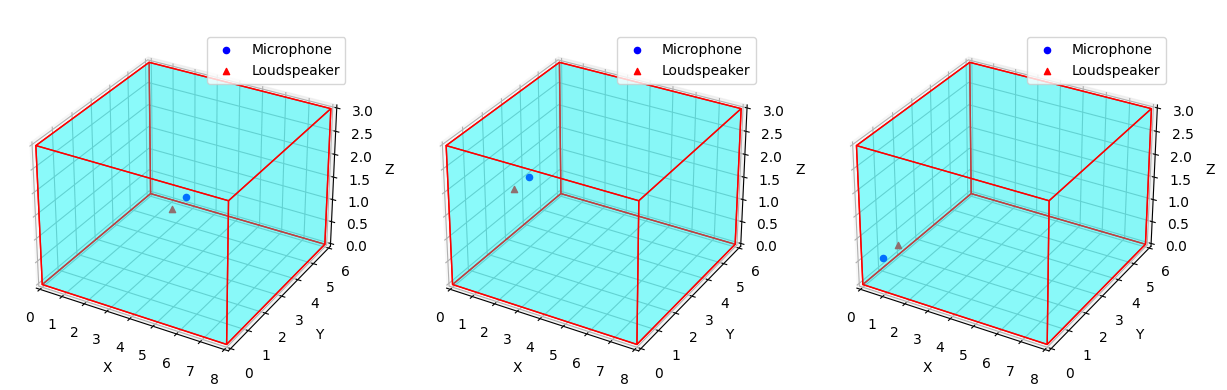}
    \caption{Room Measurement Layout Diagram}
    \label{fig:microphone position}
\end{figure}

In addition to the public datasets described above, RIRs from 11 distinct rooms at the campus of Beijing University of Technology were measured, including elevator shafts, classrooms, auditoriums, seminar rooms, and more. The parameters of these selected rooms were recorded. The aim of this endeavor is to bridge the natural gap in available real-world acoustic spaces within the volume range of 12$m^3$ to 7000$m^3$. Three RIR measurements were conducted at different positions within the selected rooms. Specifically, measurements were taken at the geometric center of the room, a location near the wall, and a position near the corner, to capture the RIR with a sequence length of 4 seconds. The microphone and loudspeaker positions are illustrated in Fig. \ref{fig:microphone position}. 

\subsection{Simulated RIR Dataset}
The real-world data is supplemented by introducing 30 simulated RIRs derived from virtual rooms with various geometries. This aims to enhance the dataset's representation of less frequently encountered room volumes, thereby achieving a normal distribution of total volume.

The specific approach involves simulating a single sound source positioned near the center of each virtual room and evenly distributing five-point receivers throughout the volume of each room. To create this synthetic dataset, the \textit{pyroomacoustics} \cite{Scheibler17} software package is deployed, which utilizes the image-source model to simulate RIRs for rooms with specific volumes. Although this geometric model does not account for phenomena like diffraction and scattering, empirical evidence demonstrates that the utilization of simulated data contributes to enhancing the model's performance, enabling it to effectively generalize to real-world data \cite{Genovese19}.

\section{Preprocessing}
In this section, we provide a detailed explanation of how audio data with reverberation and noise is generated. We started with convolving acoustic response with audio signals and adding various types of noises for subsequent neural network comparisons. 

To ensure the quality and consistency of the dataset, we performed a series of data preprocessing. Firstly, we partitioned the audio signals into training, validation, and test sets. Only real-world RIRs were selected in the test set  to asses system performance on unseen non-simulated rooms.

Furthermore, we employed a data augmentation technique called SpecAugment that aims to enhance the neural network's ability to generalize in unknown rooms and noisy environments. 

Lastly, we discussed the method for audio feature extraction. Gammatone ERB filterbank was used to generate time-frequency representations. After processing, these features resulted in a two-dimensional feature block used as input to the neural network, allowing it to handle various datasets and provide accurate blind room parameter estimation performance.

\subsection{Audio generation}

\begin{figure}
    \centering
    \includegraphics[height=8.50cm]{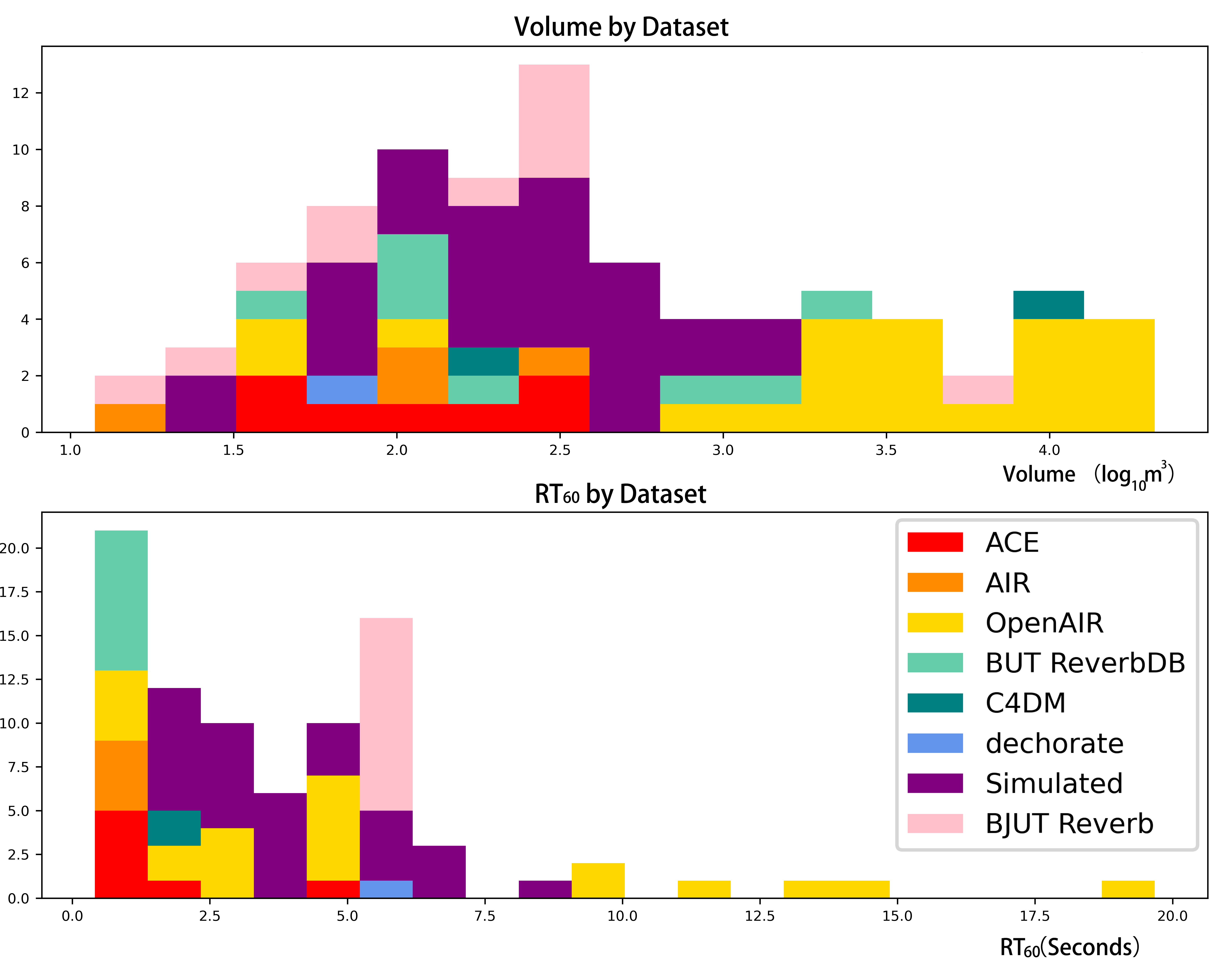}
    \caption{The distribution graphs of volume and \rts\ in different datasets.}
    \label{fig:volume rt60}
\end{figure}

In the acquired RIR dataset, a total of 55 real-world rooms and 30 simulated rooms are included, comprising a total of 570 RIRs. The volume labels span from 11.88$m^3$ to 21,000$m^3$, while the range of \rts\ varies from 0.41$s$ to 19.68$s$. Due to the significant differences in volume labels spanning multiple orders of magnitude, we chose to represent them using a logarithmic base 10 scale. Additionally, to ensure consistency across all datasets, all RIRs were downsampled to 16 kHz. The distribution of volume and \rts\ in different datasets is shown in Fig. \ref{fig:volume rt60}.

From a given RIR dataset with room parameter labels, we generated audio data with reverberation and noise for the purpose of feeding it into different neural networks for comparison. To achieve this, we mapped the acoustic response \(r(t)\) of different types of rooms in the RIR dataset onto the audio signal \(y(t)\).

We used source speech signals \(x(t)\) recorded in anechoic chambers without reverberation and convolve them with \(r(t)\) in the time domain. The source speech signals \(x(t)\) are obtained from the ACE dataset \cite{Eaton16}, where samples are recorded without reverberation, and include both male and female speakers. In the RIR dataset, some rooms have more RIR measurements than others. To ensure a uniform representation, each room is equally sampled, so that the distribution of audio samples in our dataset matches the volume distribution in our RIR dataset.

Additional noise signals \(n(t)\) were added to simulate recordings at four different signal-to-noise ratio (SNR) levels, including [+30, +20, +10, +0] decibels. The noise \(n(t)\) comprises two types of noise, namely white noise and babble noise \cite{krishnamurthy09}.

In summary, the audio signal \(y(t)\) is constructed by convolving source speech signals \(x(t)\) with room impulse response \(r(t)\) and adding additional noise \(n(t)\), represented as:

\begin{align}
y(t) &= x(t) * r(t) + n(t) \nonumber \\
     &= s(t) + n(t) \label{eq:yt}
\end{align}

Here, \(t\) represents the discrete time index. \(s(t)\) represents the reverberation audio signal without noise.

We split 32,000 audio signals \(y(t)\) into training, validation, and test sets in a 6-2-2 ratio. During the training process, we randomly sampled a subset from both real and simulated rooms for room validation. A subset was also extracted from real-world unseen rooms for room testing. Rooms with these specific parameters were not included in the training set. The purpose of this step is to assess whether the model, when confronted with room parameters not encountered during the training process, can still demonstrate robust predictive performance under noisy and reverberant conditions.
Overall, \emph{Dataset I} is formulated as listed in Table \ref{table:sum}.

\begin{table}
    \centering
    \caption{Summary of Data Splits for \emph{Datasets I} \& \emph{II}}
    \vspace{1mm}
    \begin{tabular}{c|cccc}
        \centering
        Data  &  {$\#$} of & {$\#$} of   & Real & Simulated\\
        {Split} & \emph{Dataset I}  & \emph{Dataset II} & Rooms &  Rooms\\
        \hline
        Train & 19200  & 24000 & 34 & 18 \\
        Validation & 6400  & 6400 &  21 & 12  \\
        Test & 6400 & 6400 & 21 & 0 \\
    \end{tabular}
    \vspace{-1mm}
    \vspace{-5mm}
    \label{table:sum}
\end{table}

\subsection{Audio augmentation}
\begin{figure}
    \centering
    \includegraphics[height=5.00cm]{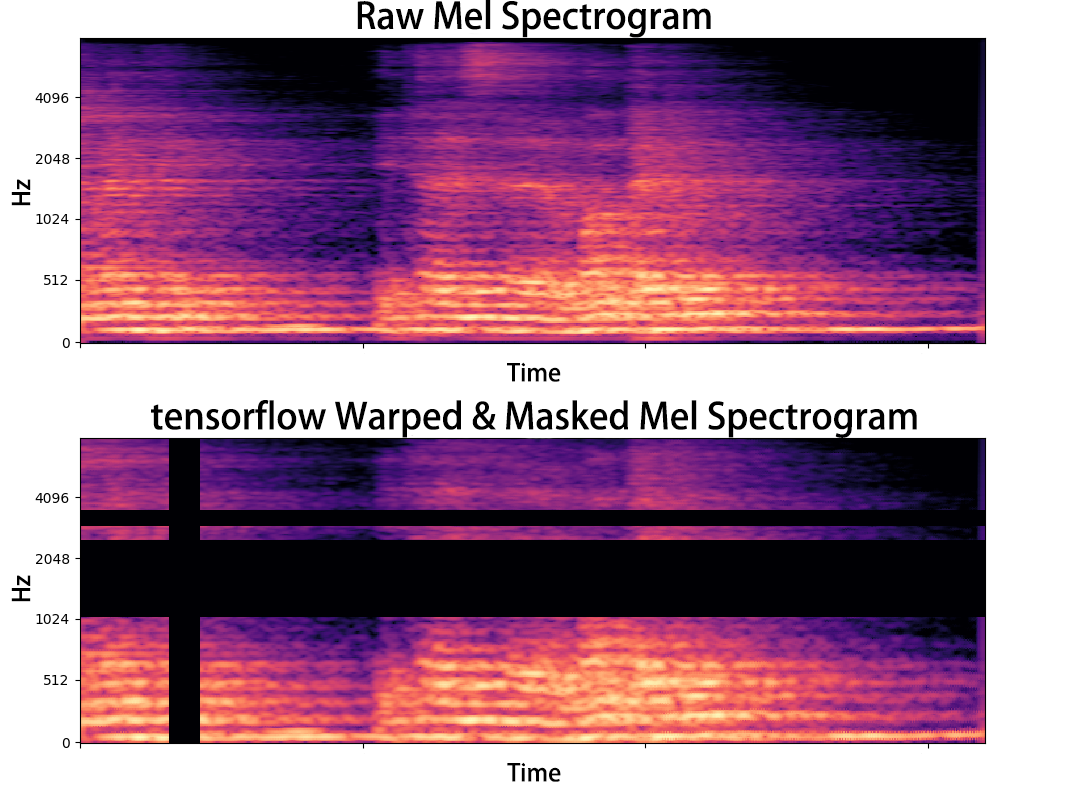}
     \vspace{3mm}
    \caption{Augmentation schemes applied to reverberation signals without noise.}
    \vspace{-1mm}
    \label{fig:specs}
\end{figure}

To enhance the generalizability of neural networks in unknown rooms and noisy environments, we employed the widely-used data augmentation technique known as SpecAugment \cite{Park}. This method enhances the model's robustness to unknown conditions through modifications and augmentations to the available training data. Specifically, we selected reverberation signals without noise \(s(t)\) as described in equation \ref{eq:yt}. Subsequently, these audio signals were transformed into log Mel-frequency spectrograms and subjected to time warping, frequency masking, and time masking, as shown in Fig. \ref{fig:specs}.

Time warping was implemented using TensorFlow's sparse image warping function. For a log Mel-frequency spectrogram with \(\tau\) time steps, we treated it as an image with the time axis horizontal and the frequency axis vertical. In the image, we randomly selected points within the interval \([W_m, \tau - W_m]\) located between time steps and applied random warping. The warping distance, \(w\), was chosen uniformly from the range \([0, W_m]\), where \( W_m\) is the time warp parameter. Six anchor points were fixed at the boundaries of the image, including the four corners and the midpoints of the vertical edges.

Frequency masking was applied as follows: a continuous set of Mel frequency channels \([f_0, f_0 + f_1)\) is masked, where \(f_1\) is initially chosen from a uniform distribution \([0, F_m]\), and \(f_0\) is chosen from \([0, \nu - f_1)\), where \( F_m\) is the frequency mask parameter, and \(\nu\) is the number of Mel frequency channels.

Time masking was applied in a similar manner: a continuous set of time steps \([t_0, t_0 + t_1)\) is masked, with \(t_1\) being initially chosen from a uniform distribution \([0, T_m]\), and \(t_0\) chosen from \([0, \tau - t_1)\), where \( T_m\) is the time mask parameter, and \(\tau\) is the total number of time steps. The Mel-frequency spectrogram with masking applied is then converted back to the time-domain signal.

Finally, 4800 speech sequences with these masking effects were added to the original training dataset for neural network training, and this dataset is labeled as \emph{Dataset II}, as shown in Table \ref{table:sum}. Constrained by computational resources, it is important to note that SpecAugment is not applied on-the-fly during each epoch. Instead, it undergoes offline processing on the data and is subsequently integrated directly into the training set. This approach aims to strike a balance between computational costs and the effectiveness of data augmentation. This comprehensive data augmentation strategy aims to help the neural network better adapt to various environments and conditions, ultimately improving its generalization performance.

\subsection{Featurization}

Audio feature extraction is crucial in convolutional neural networks, as it directly influences the model's performance. However, combining multiple feature extraction methods into one model led to complex models and requires a substantial amount of data and expensive training costs. Therefore, it is necessary to balance the addition of feature extraction methods while retaining key acoustic information to ensure that the model can handle a variety of datasets and provide general and accurate blind room parameter estimation performance.

Prior works in \cite{Genovese19,Gamper18,Srivastava22} emphasize the importance of low-frequency information for room acoustic parameter estimation. Consequently, feature representation is restricted to the relatively low-frequency range ($<$2000 kHz). The Gammatone ERB filterbank is used to generate time-frequency representations, comprising 20 frequency bands covering the frequency range from 50 Hz to 2000 Hz. The audio is computed using a 64-sample Hann window with a 32-sample hop size, resulting in a $20 \times 1997$ complex Gammatone spectrogram.

Furthermore, the phase information extracted from the audio is also retained following the work in \cite{Ick23}. Phase angles are computed for each time-frequency bin to generate phase features. These features are then truncated to include only the frequency bands associated with frequencies below 500 Hz (i.e., $5 \times 1997$) since lower-frequency components generally carry more information related to room volume. Additionally, the first-order derivatives of the phase coefficients along the frequency axis are calculated (i.e., $5 \times 1997$). This feature configuration aligns with the  `` +\emph{phase}" model described in \cite{Ick23}, which has been proven to outperform methods based solely on amplitude spectrogram features.

By combining spectral features, phase features, and first-order derivatives of phase coefficients, a two-dimensional feature block is obtained. The dimension of the feature block is $30 \times 1997$, where 30 represents the feature dimension ($F$), and 1997 represents the time dimension ($T$).

\section{Model architecture}
In this section, different architectures for audio data processing models are described for blind room parameter estimation tasks. These models include a CNN-based model, a CRNN-based model, and proposed attention-based systems.

Firstly, the CNN-based model utilizes multiple convolution and pooling layers to capture features of the audio data through convolution operations, followed by reducing the parameter count using pooling operations. Secondly, the CRNN-based model combines CNN and LSTM networks, designed to handle time series data better, capturing both time and frequency features of the audio data. Finally, the proposed model employs a completely different approach, relying solely on attention mechanisms without using convolution. This model has a unique structure, breaking inputs into patches, processing data through embedding layers and positional encoding layers, ultimately extracting features and producing results using Transformers. Additionally, the study also employs transfer learning by utilizing pretrained image models to process audio data, improving performance and efficiency.


\subsection{Convolutional neural network}

\begin{figure*}
    \centering
    \includegraphics[height=3.00cm]{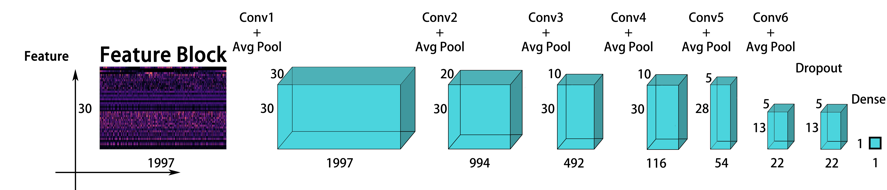}
    \caption{The system architecture of the CNN-based model.}
    \label{fig:cnn}
\end{figure*}

In this section, a model based on a CNN following the ``+\emph{phase}" model in \cite{Ick23} is introduced, for processing two-dimensional feature blocks extracted from audio data. The model comprises six convolutional layers with corresponding average pooling layers, and each convolutional layer is followed by a Rectified Linear Unit (ReLU) activation function. To prevent overfitting, dropout layers, which discard 50\% of the connections, are introduced within the network structure. Taking the estimation of room volume parameters as an example, the final output layer is a fully connected layer, mapping the output dimension to downstream tasks. In particular, the structure of its last layer is dynamically adjusted according to the requirements of the blind room parameter estimation task to meet the performance needs of different tasks. Its system architecture is illustrated in Fig. \ref{fig:cnn}.

\subsection{Convolutioanl recurrent neural network}
CRNN is designed to capture both temporal and frequency features in audio data while also having a memory to handle time series data. CRNN efficiently extracts features from data and models sequences, better accommodating variable-length inputs, making it highly suitable for practical blind room parameter estimation problems.

\begin{figure*}
    \centering
    \includegraphics[height=2.50cm]{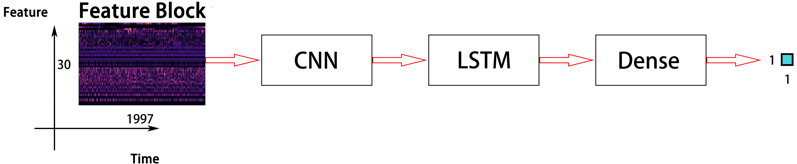}
    \caption{The system architecture of the CRNN-based model.}
    \label{fig:crnn}
\end{figure*}

A CRNN-based model is introduced in this section as it combines the parameteric efficiency of CNNs with the capability of sequential modelling from gated RNNs. The system architecture of CRNN is illustrated in Fig. \ref{fig:crnn}. 

Overall, the model consists of six convolutional layers with ReLU activation, followed by an LSTM layer, a max-pooling layer, a dropout layer, and a time-distributed fully connected layer.

Convolutional blocks gradually reduce the size and dimensions of the feature maps, allowing more sequences to enter the network. Simultaneously, the max-pooling layer is applied to 
extract useful features from the raw audio data.

The LSTM layer, which serves as a key component of the CRNN, is used for processing time series data, capturing the temporal relationships and sequence information in the input data. The hidden layer size of the LSTM is set to 64, and can be adjusted based on the size of the input fed into the LSTM. Prior to the dense layer, a max-pooling layer is employed to reduce the parameter count, with a pooling size set to 2, similar to \cite{stoter2018countnet}. 

Subsequently, the model includes a dropout layer with rate of 0.5 before the dense layer. The data is then flattened and passed to the fully connected layer, whose output size can be adjusted as needed. Considering that the estimated room parameters are positive values, an additional ReLU activation function is added to the final output layer.

Finally, the model outputs the estimated room parameters from the last time step.  In this example, blind indoor volume estimation is used as the task, which is a regression task with an output size of 1. The structure can be adapted to specific application scenarios and datasets.

\subsection{In-depth: convolution-free audio spectrogram transformer}

\subsubsection{Audio spectrogram transformer}

In this section, we introduce a model based purely on attention mechanisms without convolution for blind room parameter estimation. The design of this model is inspired by the workings of the Audio Spectrogram Transformer described in \cite{Gong21ast}, which has shown remarkable performance in end-to-end audio classification tasks. However, it is noted that this purely attention-based approach has not been extensively explored in other domains, especially in the realm of blind room parameter estimation. 

The primary goal of this section is to apply the proposed model that is purely attention-based to the blind room parameter estimation problem and compare its performance with traditional CNN and CRNN models. 

In this work, two-dimensional feature block with dimensions of $30\times1997$ as input for the proposed model is used. To better capture local information in the audio, the two-dimensional feature block is divided into $P$ patches, each sized $16\times16$. The goal of patch split is to ensure a better capture of local features within the audio signal. Additionally, to maintain consistency in both feature and time dimensions, each patch has a 6-feature dimension and 6-time dimension overlap with its surrounding blocks. As a result, the number of patches $P$ is determined to be 398, shown as:
\begin{align}
{P = \bigl\lceil \displaystyle \frac{F-16}{10} \bigr\rceil \bigl\lceil \displaystyle \frac{T-16}{10} \bigr\rceil},
\end{align}
where $F$ represents the feature dimension, and $T$ represents the time dimension.

To further process these patches, we introduced a linear projection layer. This layer's role is to flatten each $16\times16$ patch into a one-dimensional patch embedding with a dimension of $768$, referred to as the patch embedding layer. This embedding process helps reduce the data's dimensionality, making it more suitable for subsequent processing in the model.

Since these patches are not arranged in chronological order, and traditional Transformer architectures do not directly handle input sequences, we introduced trainable positional embeddings of dimension 768 in each patch. By introducing these trainable positional embeddings, the model is better able to understand the spatial structure of the audio spectrogram and grasp the positional information between patches.

Furthermore, the feature sequence is fed into the Transformer. Similar to \cite{Gong21ast}, each feature sequence begins with a \text{[CLS]} token. In this model, the encoding and feature extraction of the input sequence only utilizes the encoder part of the original Transformer architecture \cite{AV17}. The advantage of using the original Transformer structure is that it is a standard architecture already available in PyTorch and TensorFlow, making it easy to reproduce. Secondly, we plan to apply transfer learning to this task, and the standard architecture facilitates transfer learning. Specifically, the embedding size of the Transformer encoder we use is 768, with 12 layers and 12 heads, which are the same as those in \cite{dosovitskiy2020image,touvron2021training}.

\begin{figure}
    \centering
    \includegraphics[height=7.50cm]{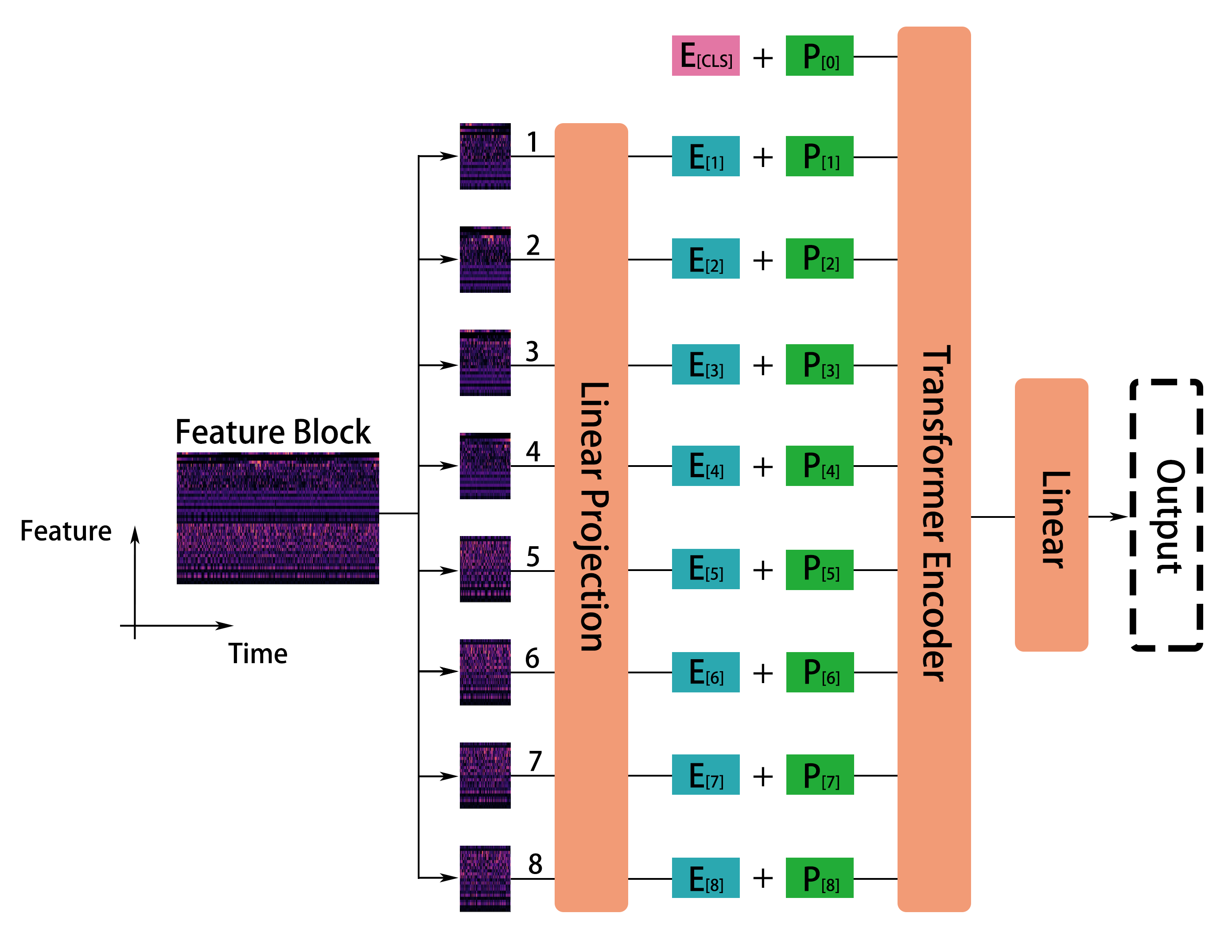}
    \caption{The system architecture of the proposed model.}
    \label{fig:ast}
\end{figure}

We adjusted the output of the encoder based on the type of room parameters being estimated. Taking room volume estimation as an example, the input consists of a sequence formed by a feature block with the dimensions of $30\times1997$, and the output is a single label used for volume prediction. The entire output of the Transformer serves as the feature representation for the two-dimensional audio feature block, which is subsequently mapped to labels for volume estimation using a linear layer with a Sigmoid activation function. The system architecture of the proposed model is depicted in Fig. \ref{fig:ast}.

In summary, traditional convolutional neural networks typically have multiple layers, small kernels, and stride sizes. In contrast to the proposed model, which includes patch embeddings (viewed as a single convolutional layer with large kernels and strides) and projection layers within Transformer blocks (viewed as $1\times1$ convolution). The proposed model used in this study can be referred to as a convolution-free model, distinguishing it from CNN and CRNN \cite{dosovitskiy2020image,touvron2021training}.

\subsubsection{ ImageNet pretraining}

Many researchers have noted that Transformers lack some of the inductive biases inherent to CNNs, such as translation equivariance and locality \cite{sun2021emergency}. Consequently, they exhibit poorer generalization capabilities under conditions of limited training data, compared to simpler models like CNNs and CRNNs \cite{dosovitskiy2020image}.

To achieve accurate blind room parameter estimation, a substantial amount of publicly available data with correctly labeled room parameters is required to train the network. Therefore, two approaches are adopted:

1) As introduced in Section 2.2, the use of an image-source model to synthesize RIR datasets.

2) Transfer learning.

Transfer learning has been widely explored in previous research, particularly in transferring from visual tasks to audio tasks. This transfer often focuses on using CNN-based models \cite{Gong21,gwardys2014deep,guzhov2021esresnet,he2019rethinking}, where ImageNet-pretrained CNN weights are used as initial weights for audio classification tasks. However, the computational cost of training state-of-the-art visual models could be relatively high. Fortunately, some common architectures like ResNet \cite{he2019rethinking} and EfficientNet \cite{tan2019efficientnet} offer readily available ImageNet-pretrained models for TensorFlow and PyTorch, making transfer learning more convenient.

For image classification tasks, research indicates that Transformer models start to outperform traditional models in performance when the dataset size exceeds 14 million \cite{dosovitskiy2020image}. However, audio datasets for blind room parameter estimation tasks typically cannot provide such massive amounts of data, posing a challenge. Therefore, we have decided to explore cross-modality transfer learning for the task of audio spectrogram processing, leveraging the similar format between image and audio data.

In this study, we use a pretrained off-the-shelf Vision Transformer (ViT) model from ImageNet \cite{touvron2021training} to simplify the transfer learning process. Afterward, we make appropriate adjustments to adapt it to the blind room parameter estimation task. Although both ViT and the proposed model employ the standard Transformer with the same patch and embedding sizes, their architectural similarities require adjustments to the structure before migration to ensure compatibility with the blind room parameter estimation task. Three adjustments are implemented:

1) One challenge is that the proposed model's input is a single-channel feature block, whereas ViT's input is a three-channel image. To overcome this issue, we adopted an approach to calculate the average weights corresponding to each of the three input channels of the ViT patch embedding layer. This averaging method helps integrate the information from the three channels into a single channel. We then used these averaged weights for the patch embedding layer. This essentially extends the single-channel spectrogram to a three-channel image with the same content but higher computational efficiency. This approach helps us better adapt to the differences in model input, thus improving the efficiency and performance of the research.

2) Another issue encountered in this study is that the input dimensions of ViT are fixed, whereas in practical tasks, the model needs to adapt to variable-length audio inputs. As the audio length changes, the dimensions of the feature block also change. While the Transformer naturally supports variable input lengths and can be directly transferred from ViT to the proposed model, special handling of positional encodings was required. This is because the ViT model learns to encode spatial information during ImageNet training. Therefore, we used the `Cut and bi-linear interpolate' method \cite{Gong21} to adjust the input size and manage positional encodings. This way, even with different input shapes, we can pass on the two-dimensional spatial knowledge obtained from the pretrained ViT to the proposed method, allowing the model to adapt to audio inputs of varying lengths. This method helps us better handle data under different input conditions.

3) To adapt to different sub-tasks in blind room parameter estimation, we take the example of blind room volume estimation. We reinitialize the final classification layer of ViT to output corresponding volume labels in the proposed model.

These adjustments are crucial to ensuring that the pretrained ViT model can be effectively used for the specific task of blind room parameter estimation and achieve improved efficiency and performance.

\section{Experiment}

In this section, an in-depth exploration of blind room parameter estimation tasks is conducted, utilizing two different datasets (\emph{Dataset I} and \emph{Dataset II}) to assess the performance of various models. We employed log-scaled and normalized data to better handle the magnitude differences between room parameters and utilized multiple evaluation metrics to comprehensively assess the accuracy and robustness of the models.

We employed various model architectures (CNN-based model, CRNN-based model, and the purely attention-based method in Section 4.3) and conducted a detailed comparison of their performance on different tasks and datasets.


Overall, through these experiments, we examined the performance of different models in various blind room parameter estimation tasks and assessed their adaptability in handling variable-length audio inputs. 

\subsection{Datasets}
In the task of blind room parameter estimation, \emph{Dataset I} and \emph{Dataset II} mentioned in sections 3.1 and 3.2 were utilized. In the preprocessing phase, room volume labels (in m$^3$ units, in logarithmic scale) were exclusively read, and four models, CNN-based model, CRNN-based model, the Proposed Method, and the ``proposed method w/ pretrain" model, were individually evaluated for their performance on  \emph{Dataset I} and \emph{Dataset II}. For the blind room parameter estimation with variable-length audio input, \emph{Dataset II} was employed. Similarly, in the preprocessing phase, only room volume labels (in m$^3$ units, in log-scaled) were considered. However, a modification was made to the test set of \emph{Dataset II}. Specifically, samples were extracted from 1 to 4 seconds with a step size of 0.5 seconds, and zero padding was applied to different lengths of audio samples to match the original length. This was done to assess the performance of different models in handling blind room parameter estimation under audio inputs with different length.

Finally, in the task of joint estimation of room parameters, \emph{Dataset II} was used. In the preprocessing phase, the model simultaneously reads room \rts\ (in seconds) labels and room volume labels (in m$^3$ units). In order to overcome the significant scale differences between these two parameters, we adopted an approach where we mapped the values of \rts\ to volume values and applied a logarithmic scaling to them. It is worth emphasizing that this data processing method is reversible, allowing us to revert all parameters to standard units at any time. The advantage of mapping the parameter relationship rather than standard normalization is that it eliminates the need for frequent adjustment of hyperparameters when dealing with different blind room parameter estimation tasks, as it effectively addresses the differences in units and magnitudes among the parameters. This is done to evaluate the performance of different models in joint room parameters estimation.

\subsection{Evaluation metrics and loss function}
As shown in Fig. \ref{fig:volume rt60}, due to large span of room volume and \rts\ ranges, the estimation error could be related to its order of magnitude. Therefore, a log-10 estimation is more suitable than a linear estimation. This way, larger acoustic spaces in training are not disproportionately affected due to the relatively high contribution of error estimation. Using a logarithmic estimation better handles estimation errors of different orders of magnitude.

Four evaluation metrics using a base-10 logarithm are considered. They are as follows:
1) Mean Squared Error (MSE): MSE is the average of the squared differences between estimated room parameters and ground truth room parameters. It is used to measure the degree of dispersion between estimated values and ground truth values. The smaller the average of squared differences, the closer the estimated values are to the ground truth values.
2) Mean Absolute Error (MAE): MAE represents the average of the absolute differences between estimated values and ground truth values. It provides the average deviation between estimated values and ground truth values and is commonly used to measure the accuracy of estimated values.
3) Pearson Correlation Coefficient ($\rho$): The Pearson correlation coefficient is used to measure the strength and direction of the linear relationship between two variables. It is used to describe the relationship between estimated room parameters and ground truth room parameters, with values ranging from -1 to 1. Negative values indicate a negative correlation, positive values indicate a positive correlation, and 0 indicates no correlation.
4) MeanMult \textit{MM}: \textit{MM} is the mean absolute logarithm of the ratio between the estimated room volume and the ground truth room volume. This metric provides an overview of the mean error in the ratio between estimated room parameters and ground truth room parameters. Taking the logarithm of the ratio helps reduce the impact of data points with significant differences. For example, for the estimated volume parameter, given the estimated volume  $\hat{V}_n$ and the ground truth volume  $V_n$:
\begin{equation}
    \textit{MM} = e^{\frac{1}{N} \sum_{n=1}^{N} | \ln\left(\frac{\hat{V}_n}{V_n}\right) |},
\end{equation}
where ``$n$" represents the sample index, and ``$N$" represents the total number of samples.

During model training stages, MSE was used as the loss function to minimize the error between estimated room parameters and ground truth room parameters. In the ``Estimation of room parameter" and ``Room parameter estimation under variable-length audio input" tasks, the loss function $\mathrm{L_1}$ formula was as follows:

\begin{equation}
\text{L}_1 = \frac{1}{B} \sum_{n=1}^{B} (\hat{V}_n - V_n)^2\label{eq:l11},
\end{equation}

where ``$n$" represents the sample index, and ``$B$" represents the batch size during training.

In contrast, for the task of “Joint estimation of room parameters,” which involves the simultaneous estimation of \rts\ and volume parameters. To avoid differences in units and orders of magnitude between different parameters, as well as the impact of parameter scaling methods, the normalized MSE was used instead of the MSE in Eq. \ref{eq:l11}. The loss function $\mathrm{L_2}$ was formulated as follows:

\begin{equation}
\text{L}_2 =  {\lambda_1} * \frac{\sum_{n=1}^{B} (\hat{U}_n - U_n)^2}{B\sum_{n=1}^{B}(U_n)^2},+{\lambda_2} * \frac{\sum_{n=1}^{B} (\hat{V}_n - V_n)^2}{B\sum_{n=1}^{B}(V_n)^2}\label{eq:l12},
\end{equation}
where  $\hat{U}_n$ and $U_n$ represent the estimated and the ground truth \rts, respectively. $\lambda_1$ and $\lambda_2$ are weights used to control the balance between the \rts\ and volume normalized MSE loss functions. These weights are employed to adjust the relative importance of these two functions during model training. Based on empirical evidences and experimental results, $\lambda_1$ is set to 1, and $\lambda_2$ is set to 2. This weight configuration can be adjusted according to the specific task and model performance to better meet the training requirements.

\subsection{Experiment configurations}

Different MSE loss functions were chosen based on the task's requirements. Each model utilized the Adam optimizer from PyTorch. During the training process, L2 regularization was applied to prevent overfitting. Simultaneously, an adaptive learning rate strategy was employed to ensure the convergence of the model. If the model's validation set did not improve for ten consecutive epochs, an early stopping criterion was triggered, leading to the cessation of the training process to prevent further overfitting. Furthermore, to select the optimal-performing model, we monitored the MSE values on the validation set during grid search and optimized hyperparameters, including initial learning rate as well as batch size. The hyperparameter configuration that demonstrated the best performance was chosen as the final model parameters.

For the ``Estimation of room parameter" task, to facilitate comparative testing, we switched between \emph{Dataset I}  and \emph{Dataset II} as well as determined whether to use a pretrained model from ImageNet. To ensure consistency in model configurations, hyperparameters were kept constant. CNN-based and CRNN-based models were trained for 1000 epochs with an initial learning rate of 5e-4, a batch size of 128.

The proposed attention-based method and the ``proposed method w/ pretrain" model were trained for 150 epochs with an initial learning rate of 5e-6, a batch size of 16. For the ``Joint estimation of room parameters" task, CNN-based and CRNN-based models were trained for 2000 epochs with an initial learning rate of 2e-4, a batch size of 128. The proposed method and the ``proposed method w/ pretrain" model were trained for 300 epochs with an initial learning rate of 2e-6, a batch size of 16.

To ensure fairness, all models were trained on devices equipped with an Intel Core i9 processor and an NVIDIA GeForce 4090 GPU.

\section{Results and discussion}

\subsection{Estimation of room volume parameter}

To investigate whether comparable performance similar to that of CNN and CRNN can be achieved by using a pure attention mechanism, we extracted audio data from \emph{Dataset I} for the purpose of estimating room volume parameter. We transformed audio data into a feature block, as described in Section 3.3. Subsequently, we separately input these feature blocks into the CNN-based model, CRNN-based model, and the proposed method (the base version without ImageNet pretraining) for training. We then compared the predicted volume labels to the ground truth values. The results of these three models are presented in Table \ref{table:per1}. Note that in this section we mainly focus on estimation of room volumes as this task has been shown to be more challenging than \rts\ estimation in the literature \cite{Gamper18,Genovese19,Ick23}.

In addition, the table includes information on the model's memory consumption and computational complexity, such as the number of parameters ($\#$Param) and Multiply–Accumulate Operations (MACs). To ensure fairness, we conduct tests using the PyTorch profiler \cite{pytorch_profiler_tutorial} in the same GPU environment. A comprehensive comparison of the data in the table reveals that CNN models have fewer parameters and relatively low memory consumption but perform worse in various evaluation metrics. While the CRNN model shows an increase in parameter count compared to CNN and some improvement in evaluation metrics, it still falls short of our proposed method. In contrast, although our method has relatively higher parameter count and memory consumption, it demonstrates significant advantages in all evaluation metrics. Specifically, our method outperforms in terms of MSE, MAE, $\rho$, \textit{MM}, and MACs. Despite the relatively higher memory consumption, considering the performance improvement, this increase can be deemed acceptable.

\begin{table}
\caption{The comparison between the CNN-based model \cite{Ick23}, the CRNN-based model, and the base version of the proposed method.}
\centering
\begin{tabular}{c|ccccccc}
\multirow{3}{*}{Method} & \multirow{2}{*}{$\#$ Params} &\multicolumn{4}{c}{Evaluation Metrics} &{Memory} &\multirow{2}{*}{MACs} \\
\cmidrule{3-6}
& (M) & \multirow{2}{*}{MSE} & \multirow{2}{*}{MAE} & \multirow{2}{*}{$\rho$} & \multirow{2}{*}{\textit{MM}} & Consumption & (G)\\
&  & &  &  & & (GB)) & \\
\hline
CNN \cite{Ick23} & 0.013 & 0.3863 & 0.4837 & 0.6984 & 3.0532 & 1.81 & 0.237 \\
CRNN  & 0.494 & 0.3572  & 0.4265 & 0.7262 & 2.6701 & 1.95 & 0.236\\
\textbf{Proposed} & \multirow{2}{*}{85.256} & \multirow{2}{*}{\textbf{0.2650}}  & \multirow{2}{*}{\textbf{0.3432}} &  \multirow{2}{*}{\textbf{0.8077}} & \multirow{2}{*}{\textbf{2.2039}} & \multirow{2}{*}{{4.55}} & \multirow{2}{*}{{34.083}} \\
\textbf{method} & & & & & & & \\
\end{tabular}

\vspace{3mm}
\label{table:per1}
\end{table}

Based on experimental results above, we can see that the proposed method using the pure attention mechanism significantly outperforms both CNN and CRNN-based approaches, even with a lower-layer network configuration and relatively fewer training epochs. This suggests that the proposed method can accurately capture the acoustic characteristics in the audio data, thereby improving the accuracy and stability of room volume estimation.

Meanwhile, the four evaluation metrics show that the CRNN-based model performs better than the CNN-based model. This can be attributed to the advantages of CRNN, which combines CNN with LSTM. CRNN can better handle the time series audio data while capturing local features, which is crucial for blind room parameter estimation tasks.


\begin{table*}[h]
\caption{ Performance comparison of different models with and without the application of SpecAugment.}
\centering
\begin{tabular}{c|c c c p{0.80cm}|c c c p{0.80cm}}
\hline
\multirow{2}{*}{Method} & \multicolumn{4}{c|}{\emph{Dataset I}} & \multicolumn{4}{c}{\emph{Dataset II}} \\
\cmidrule{2-9}
&MSE &MAE & $\rho$ & \textit{MM} & MSE &MAE & $\rho$ & \textit{MM} \\
\hline
CNN \cite{Ick23} & 0.3863 & 0.4837 & 0.6984 & 3.0532 & 0.3154 & 0.4136 & 0.7678 & 2.5921 \\
\hline
CRNN  & 0.3572 & 0.4265 & 0.7262 & 2.6701 & 0.2818 & 0.3684 & 0.7898 & 2.3471\\
\hline
Proposed &\multirow{2}{*}{0.2650} & \multirow{2}{*}{0.3432} & \multirow{2}{*}{0.8077} & \multirow{2}{*}{2.2039} & \multirow{2}{*}{0.1981} & \multirow{2}{*}{0.2884} & \multirow{2}{*}{0.8580} & \multirow{2}{*}{1.9427} \\
\textbf{method}  & & & & & &\\
\hline
\textbf{Proposed}  & \multirow{3}{*}{0.2157} & \multirow{3}{*}{0.3111} & \multirow{3}{*}{0.8529} & \multirow{3}{*}{{2.0470}} & \multirow{3}{*}{\textbf{0.1541}} &\multirow{3}{*}{\textbf{0.2423}} & \multirow{3}{*}{\textbf{0.8929}} & \multirow{3}{*}{\textbf{1.7470}} \\
\textbf{method}  & & & & & &\\
\textbf{w/pretrain}  & & & & & &\\
\hline
\end{tabular}
\vspace{-1mm}
\vspace{-1mm}
\label{table:spec}
\end{table*}

To further investigate the impact of ImageNet pretraining on the proposed method's performance, the ``proposed method w/ pretrain" model was trained on \emph{Dataset I}. Simultaneously, to examine the effect of the SpecAugment data augmentation method on the performance of existing models, we retrained the existing four models on \emph{Dataset II}. The results of the above experiments are shown in Table \ref{table:spec}.

Based on the training results of different models on \emph{Dataset I}, we can observe a significant improvement in the performance of the proposed method in the ``Estimation of room parameter" task with the use of the ImageNet pretraining method. Furthermore, when the four models were retrained on \emph{Dataset II}, the application of the SpecAugment method elevated the models' performance to a new level. In particular, this method demonstrates a significant improvement in the performance of the ``proposed method w/ pretrain" model. It confirms the effectiveness of SpecAugment in mitigating overfitting and enhancing model generalizability.

Meanwhile, in order to provide a more illustrative example, we rescaled the experimental results to a linear scale. In this experiment, the test set room volume ranges from 12 to 21,000 $m^3$. We compared the performance of the best-performing models, namely the CNN-based model, CRNN-based model,  and the ``proposed method w/ pretrain" model. They were trained on \emph{Dataset II}, and their model's median as well as mean absolute error are shown in Table \ref{table:paa}.

\begin{table}
\caption{Comparison of median and mean absolute error for volume parameters among the best-performing models.}
\centering
\begin{tabular}{c|cc}
\multirow{2}{*}{Method} & Median & Mean absolute error  \\
\ & ($m^3$)& $(m^3$)  \\
\hline
CNN \cite{Ick23} & 353 & 1919 \\
CRNN  & 257  & 1644 \\
\textbf{Proposed method}  & \textbf{\multirow{2}{*}{155}} & \textbf{\multirow{2}{*}{1219}}\\
\textbf{w/pretrain}  & & \\
\end{tabular}
\vspace{3mm}
\label{table:paa}
\end{table}

From the table, the ``proposed method w/ pretrain" model exhibits the best performance in terms of both median and mean absolute error, having the lowest error values.

\begin{figure}[h]  
    \centering
    \begin{subfigure}{0.25\textwidth}
        \captionsetup{justification=centering} 
        \centering
        \includegraphics[height=3.00cm]{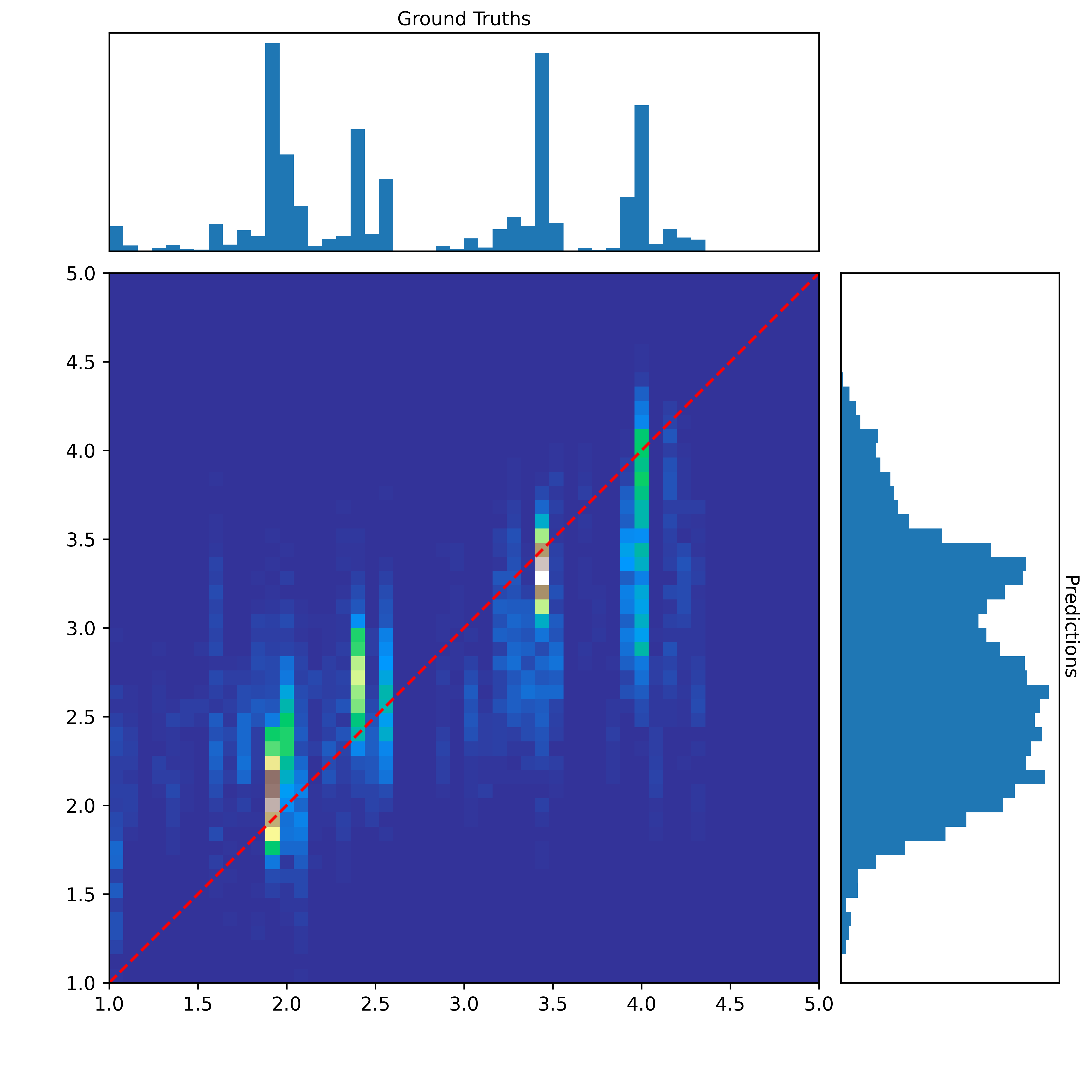}
        \caption{CNN-based model}
        \label{fig:subfig1}
        \vspace{8mm} 
    \end{subfigure}
    \hfill
    \begin{subfigure}{0.25\textwidth}
        \captionsetup{justification=centering} 
        \centering
        \includegraphics[height=3.00cm]{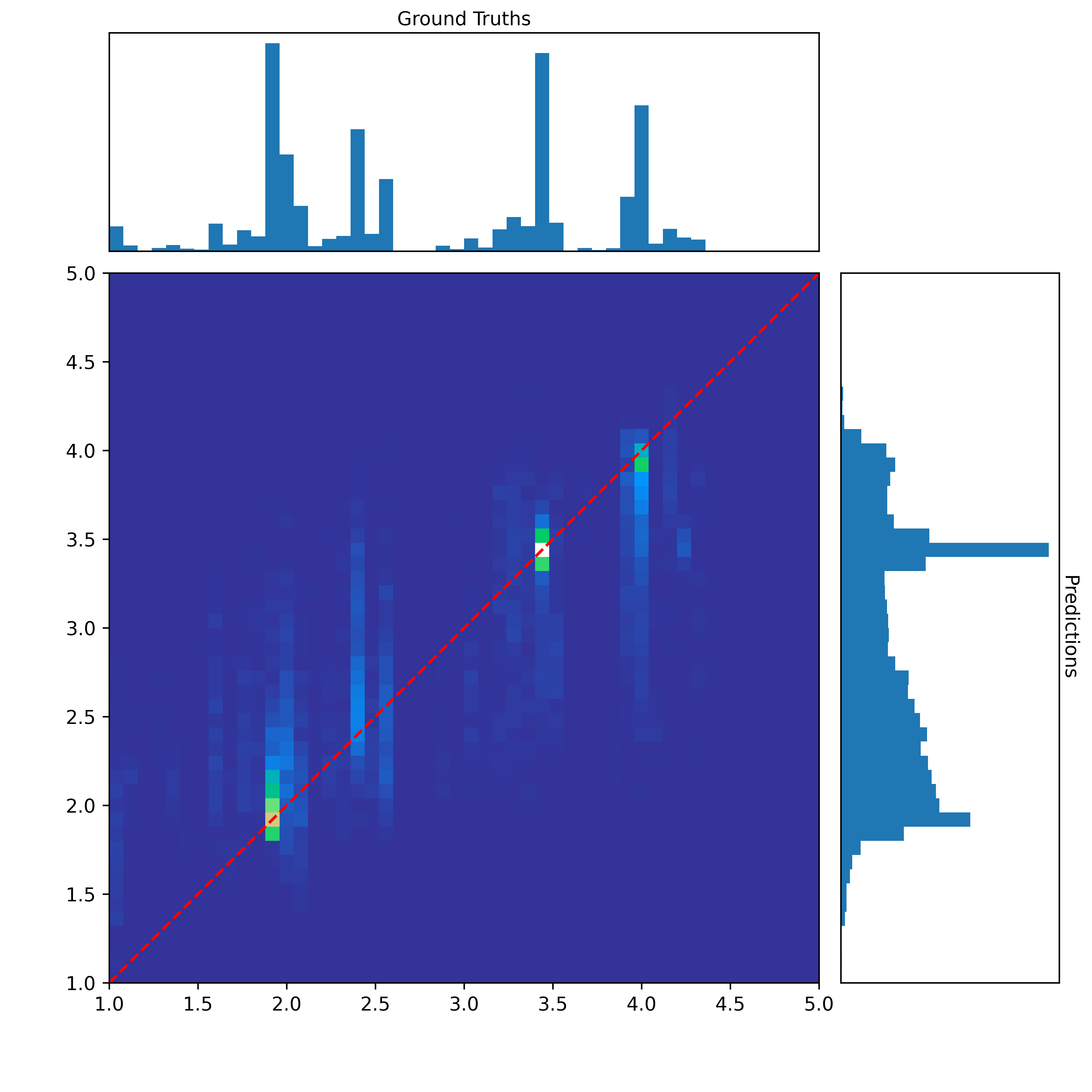}
        \caption{CRNN-based model}
        \label{fig:subfig2}
       \vspace{4mm} 
    \end{subfigure}
    \hfill
    \begin{subfigure}{0.25\textwidth}
        \captionsetup{justification=centering} 
        \centering
        \includegraphics[height=3.00cm]{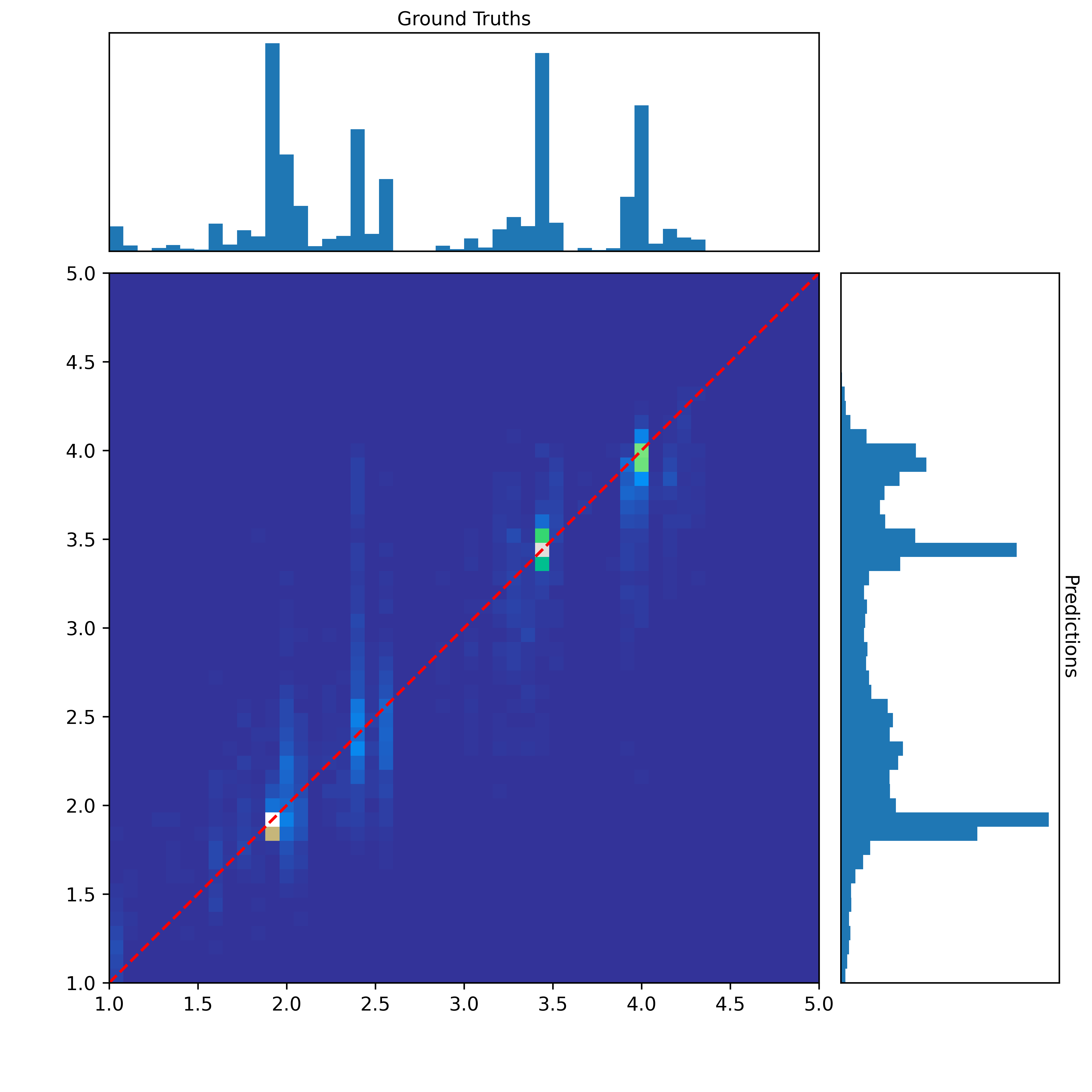}
        \caption{The ``proposed method w/ pretrain" model}
        \label{fig:subfig3}
    \end{subfigure}
    \caption{Confusion Matrices for the best-performing models trained on \emph{Dataset II} in the ``Estimation of room volume parameter" task}
    \label{fig:volhots}
\end{figure}

The Fig. \ref{fig:volhots} displays the confusion matrices for these three best-performing models in the ``Estimation of room volume parameter" task, with the x-axis and y-axis representing the log-10 exponent of volume size. From the visualization, it is evident that the ``proposed method w/ pretrain" model exhibits excellent performance across the entire test range. Its distribution consistently closely surrounds the ground truth, clearly outperforming the CNN-based and CRNN-based models.

Results in this section indicate that the proposed purely attention-based model is capable of capturing relevant features and representations in the context of room volume regression efficiency. More importantly, it demonstrates remarkable generalization capabilities, effectively applying the patterns learned from the training data to real-world rooms, even for rooms the model has not encountered before, resulting in accurate volume estimates. This outcome provides a strong theoretical foundation for our approach and underscores its potential in more blind estimation practical problems, which will be addressed in the following section.

\subsection{Room parameter estimation under variable-length audio input}

\begin{figure*}
    \centering
    \includegraphics[height=10.10cm]{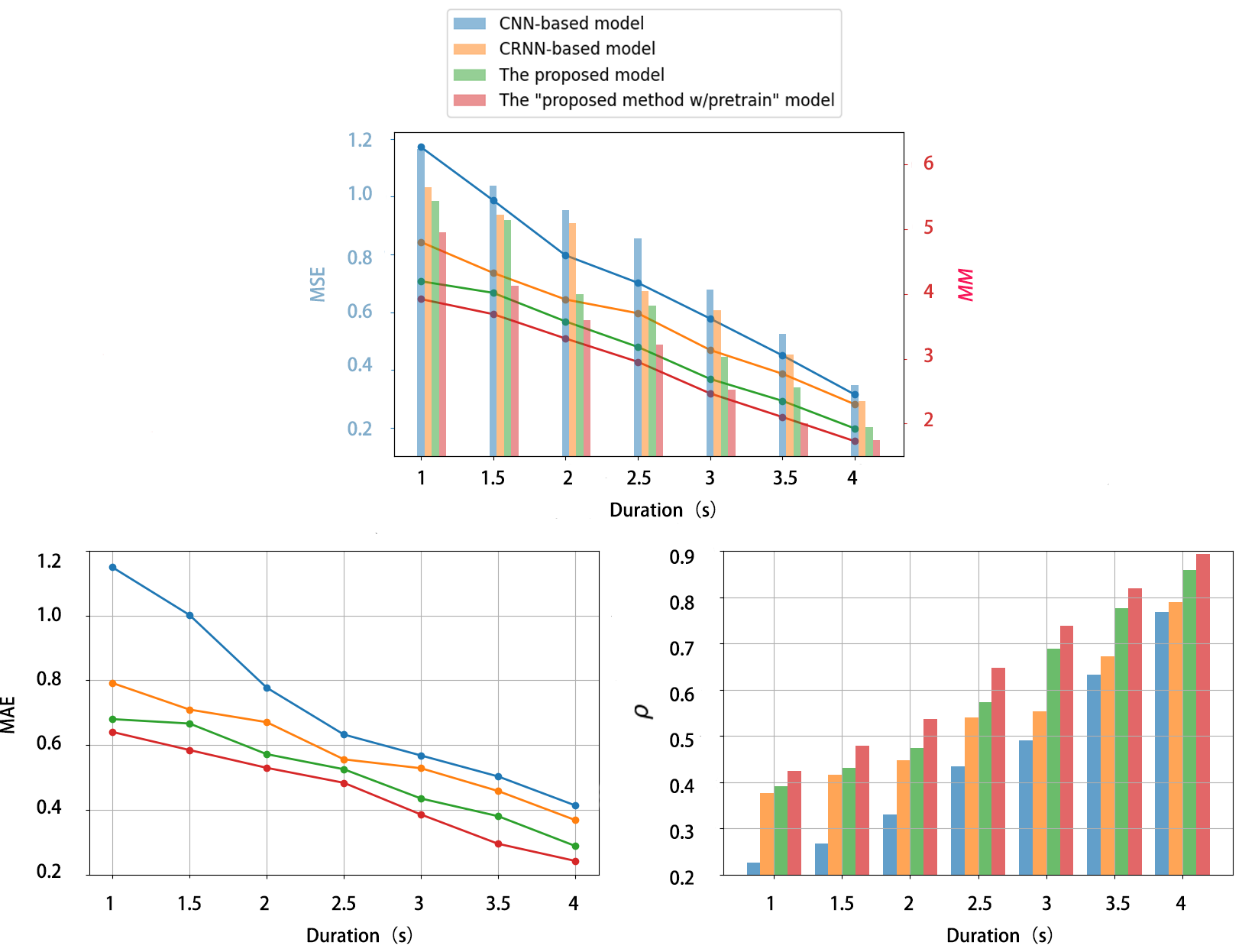}
    \captionsetup{justification=centering}
    \caption{Performance comparison of different models under the ``Room parameter estimation under variable-length audio input'' task.}
    \label{fig:duration}
\end{figure*}

In this section, model performances under variable-length audio inputs are evaluated for the ``Room parameter estimation" task. The selected models were tested with different lengths of audio inputs, and their performances were assessed using four objective evaluation metrics as shown in Fig. \ref{fig:duration}. It is evident from the figure that the accuracy of the models in predicting room volume parameter significantly depends on the length of the input audio. As the input audio length shortens, the estimation performance of all models inevitably experiences degradation.

By observing the curves of MSE and MAE metrics in Fig. \ref{fig:duration}, it can be noted that the CRNN-based model, the proposed model, and the ``proposed method w/ pretrain" model exhibit smaller decay slopes. This suggests that, compared to CNN-based models, they can better handle time sequences of variable length. The smaller decay slopes of the evaluation metrics can be considered an indication that the models better maintain performance stability, even when the input length decreases, maintaining relatively good performance.

Results in Fig. \ref{fig:duration} also indicate that the ``proposed method w/ pretrain" model performs the best at the same input length. For the shortest input sample, i.e., when the input length is 1 second, the MSE for the ``proposed method w/ pretrain" model is 0.6458. In comparison, to achieve the same performance level, the CNN-based model, CRNN-based model, and the proposed model would require input audio lengths of approximately 2.8 seconds, 2.0 seconds, and 1.2 seconds, respectively. This advantage facilitates the proposed attention-based models to outperform both CNN and CRNN systems with significantly less temporal context, which can be a valuable merit when dealing with speech-based blind estimation problems in practice.

\subsection{Joint estimation of room parameters}

This section aims to address the ``Joint estimation of room parameters" task, which involves training a single model to simultaneously estimate multiple room parameters. Specifically, due to the shared acoustic characteristics of room volume and \rts\ parameters, it is possible to estimate them concurrently by extracting reverberation-related information. Considering difficulties in collecting groundtruth of real data for other room parameters, such as total surface area and average surface absorption coefficient in real-world room datasets, this experiment focuses on the joint estimation of room volume and \rts.

In this task, we selected three models, namely the CNN-based model, CRNN-based model,  as well as the ``proposed method w/ pretrain" model, and trained them on \emph{Dataset II}. Their network architectures were fundamentally similar to those used for the ``Estimation of room volume parameter" task, with minor modifications. In the ``Joint estimation of room parameters" task, the three models are required to output two parameters, i.e. room volume and \rts, instead of a single parameter. Consequently, the final output layers of the models were modified to include two fully connected layers for estimating different room parameters. During the training process, hyperparameters were fine-tuned (as described in Section 5.3), and the loss function was adjusted (as shown in Eq.\ref{eq:l12}).

It is worth noting that, in order to mitigate issues related to different units and scales among parameters, as well as the impact of parameter scaling during normalization, we chose to use only the $\rho$ as the evaluation metric. This helps ensure consistency among the estimated parameters. The corresponding results are presented in the Table \ref{table:rho}.

\begin{table*}[h]
\caption{Pearson correlation coefficients of best-performing models in ``Joint estimation of room parameters" task.}
\centering
\begin{tabular}{c|c c |c c| c p{1cm}}
\hline
\multirow{3}{*}{Method} & \multicolumn{2}{c|}{\multirow{2}{*}{CNN \cite{Ick23}}} & \multicolumn{2}{c|}{\multirow{2}{*}{CRNN}}  & \multicolumn{2}{c}{\multirow{2}{*}{\begin{tabular}[c]{@{}c@{}}Proposed method\\ w/ pretrain\end{tabular}}} \\
& & & & & & \\\cmidrule{2-7}
&vol &\rts &vol &\rts &vol &\rts \\
\hline
$\rho$  & 0.6187 & 0.9133 & 0.6584 & 0.9488 & \textbf{0.8287} & \textbf{0.9681}  \\
\hline
\end{tabular}
\vspace{-1mm}
\vspace{-1mm}
\label{table:rho}
\end{table*}

From these results, it can be clearly seen that the ``proposed method w/ pretrain" model outperforms the other models, achieving the highest $\rho$ for both room volume and \rts, indicating its effectiveness in jointly estimating these room parameters.

In this experiment, the test set room volume ranges from 12 to 21,000 $m^3$ while the \rts\ range from 0.41 to 19.68 seconds. We rescaled the experimental results to a linear scale. The median, as well as mean absolute error for the three models regarding volume and \rts, are displayed in Table \ref{table:qws}.

Furthermore, we conducted a comparative study between the volume estimation in the joint model and the estimation of volume results for the ``Estimation of room volume parameter" task by comparing results in Table \ref{table:spec} and Table \ref{table:rho}, as well as Table \ref{table:paa} and Table \ref{table:qws}. Despite the fact that the joint estimation models aim to simultaneously handle multiple parameters, it is clear that their volume estimation results, while experiencing some degree of attenuation, are overall very similar to the results obtained from estimating only a single parameter. This suggests that the performance of the joint model is in par with that of models designed for estimating a single parameter.

\begin{table*}[h]
\caption{Comparison of median and mean absolute error for volume as well as \rts\ parameters among the best-performing models.}
\centering
\begin{tabular}{c|cc|cc}
\hline
\multirow{3}{*}{Method} & \multicolumn{2}{c|}{Median} & \multicolumn{2}{c}{Mean absolute error} \\
\cmidrule{2-5}
&vol &\rts &vol &\rts \\
&($m^3$)&(seconds)&($m^3$)&(seconds)\\
\hline
CNN \cite{Ick23}  & 728 & 0.64 & 2481 & 1.32 \\
CRNN & 329 &0.39 & 2265 & 0.71 \\
\textbf{Proposed method}  & \textbf{\multirow{2}{*}{294}} & \textbf{\multirow{2}{*}{0.31}}& \textbf{\multirow{2}{*}{2208}} & \textbf{\multirow{2}{*}{0.61}}\\
\textbf{w/pretrain}  & & \\
\hline
\end{tabular}
\vspace{-1mm}
\vspace{-1mm}
\label{table:qws}
\end{table*}

\begin{figure}[h]  
    \centering
    \begin{subfigure}{0.25\textwidth}
        \captionsetup{justification=centering} 
        \centering
        \includegraphics[height=3.00cm]{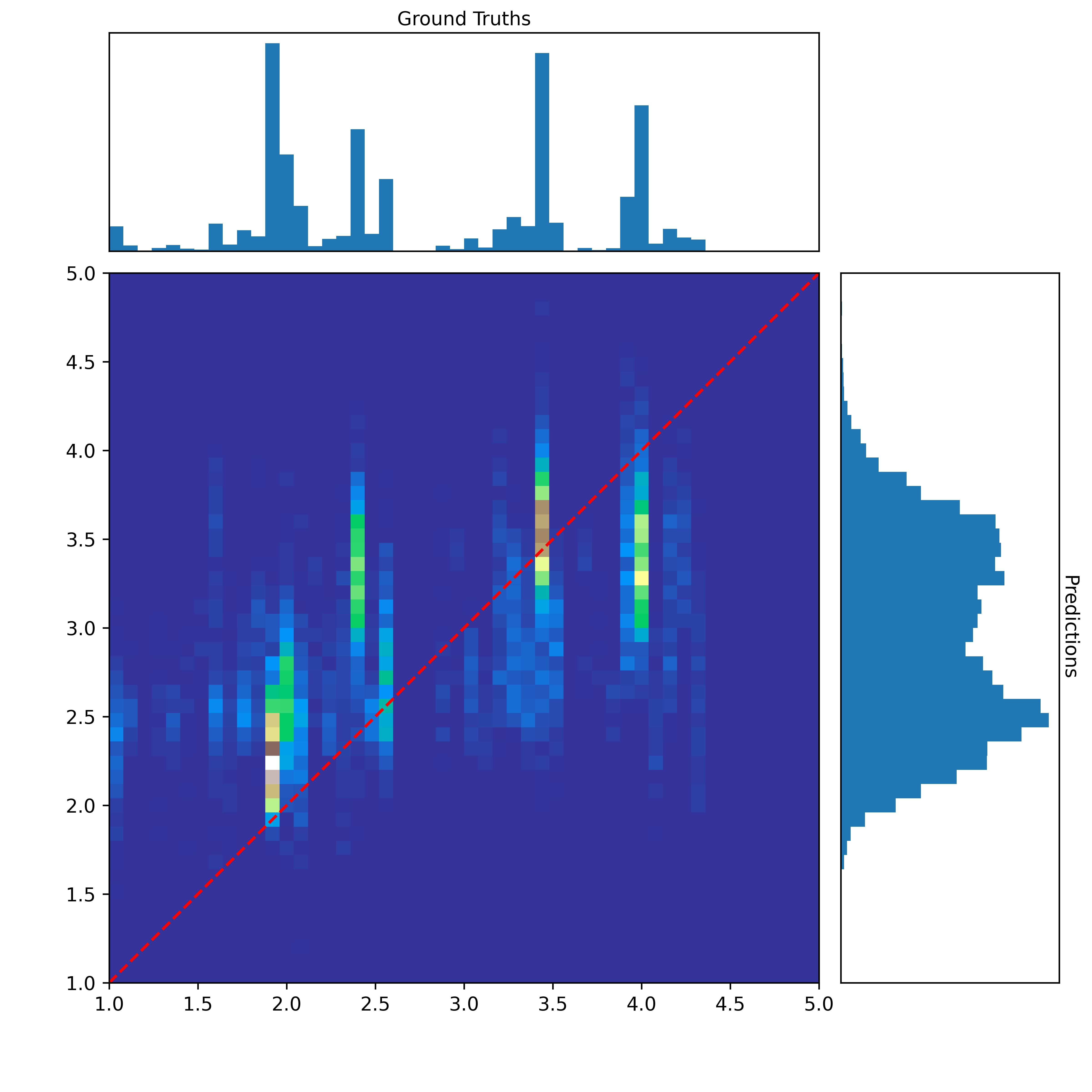}
        \caption{CNN-based model(volume)}
        \label{fig:subfig11}
        \vspace{4mm} 
    \end{subfigure}
    \hfill
    \begin{subfigure}{0.25\textwidth}
        \captionsetup{justification=centering} 
        \centering
        \includegraphics[height=3.00cm]{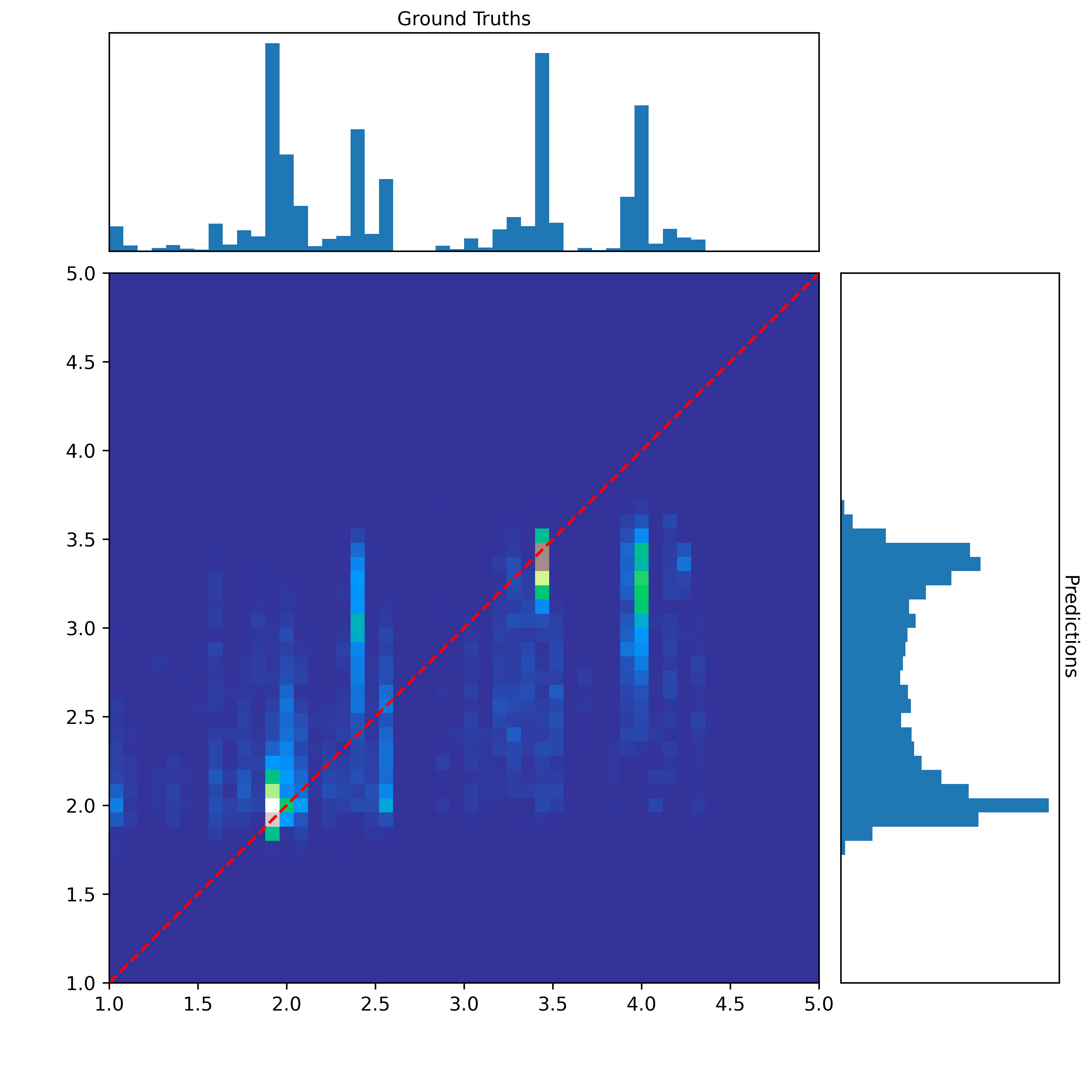}
        \caption{CRNN-based model(volume)}
        \label{fig:subfig21}
       \vspace{4mm} 
    \end{subfigure}
    \hfill
    \begin{subfigure}{0.25\textwidth}
        \captionsetup{justification=centering} 
        \centering
        \includegraphics[height=3.00cm]{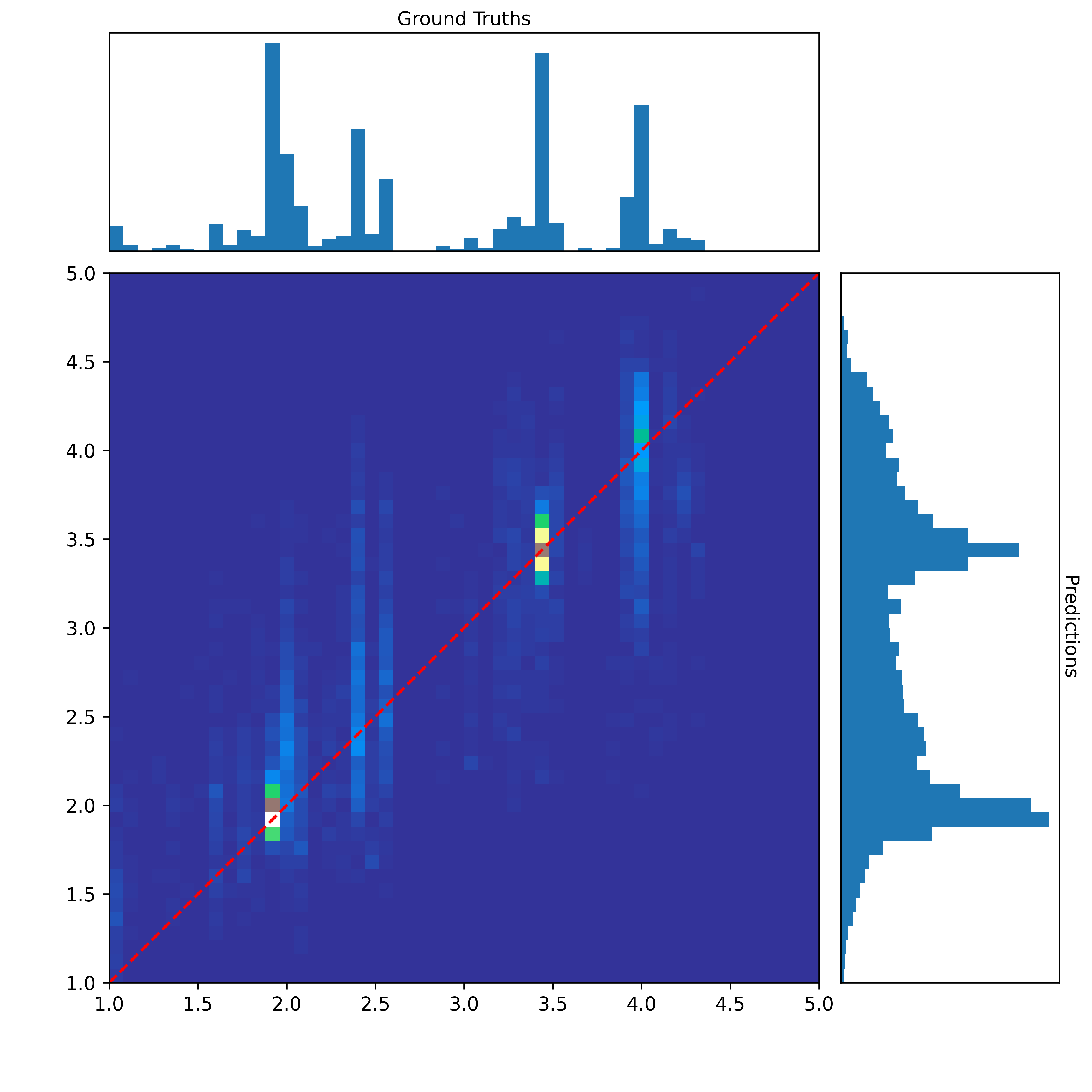}
        \caption{The ``proposed method w/ pretrain" model(volume)}
        \label{fig:subfig31}
    \end{subfigure}

    \begin{subfigure}{0.25\textwidth}
        \captionsetup{justification=centering} 
        \centering
        \includegraphics[height=3.00cm]{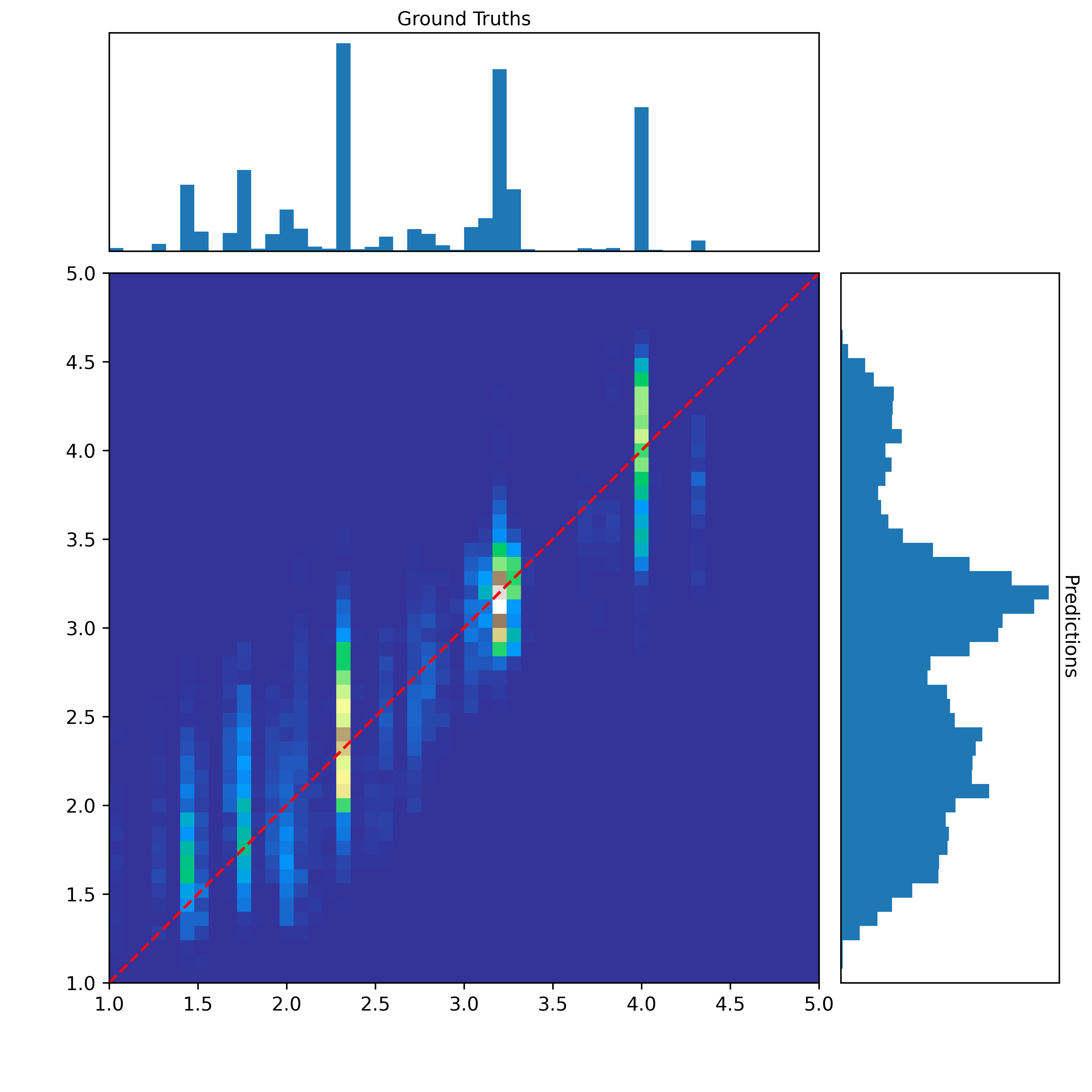}
        \caption{CNN-based model(\rts)}
        \label{fig:subfig12}
        \vspace{4mm} 
    \end{subfigure}
    \hfill
    \begin{subfigure}{0.25\textwidth}
        \captionsetup{justification=centering} 
        \centering
        \includegraphics[height=3.00cm]{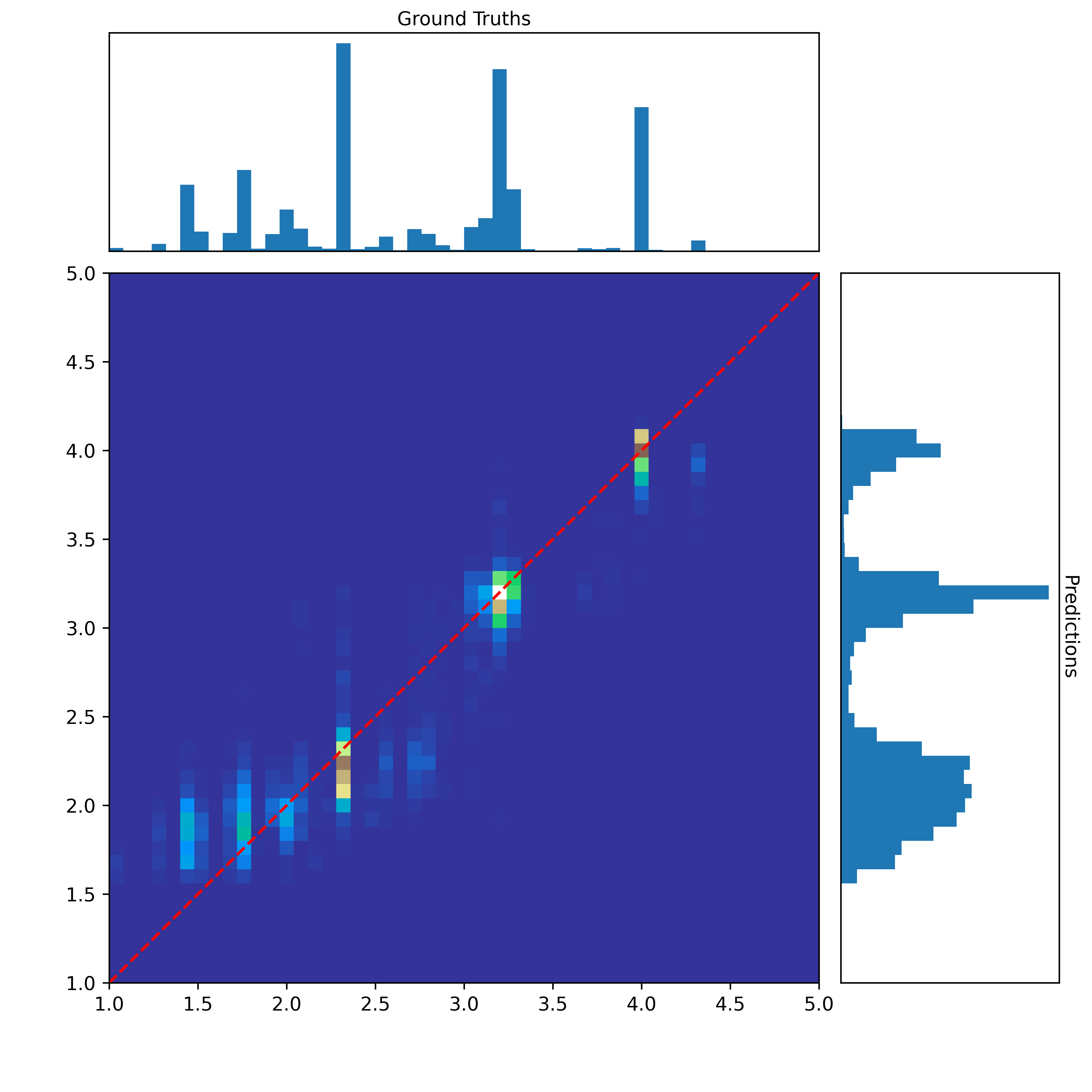}
        \caption{CRNN-based mode(\rts)}
        \label{fig:subfig22}
       \vspace{4mm} 
    \end{subfigure}
    \hfill
    \begin{subfigure}{0.25\textwidth}
        \captionsetup{justification=centering} 
        \centering
        \includegraphics[height=3.00cm]{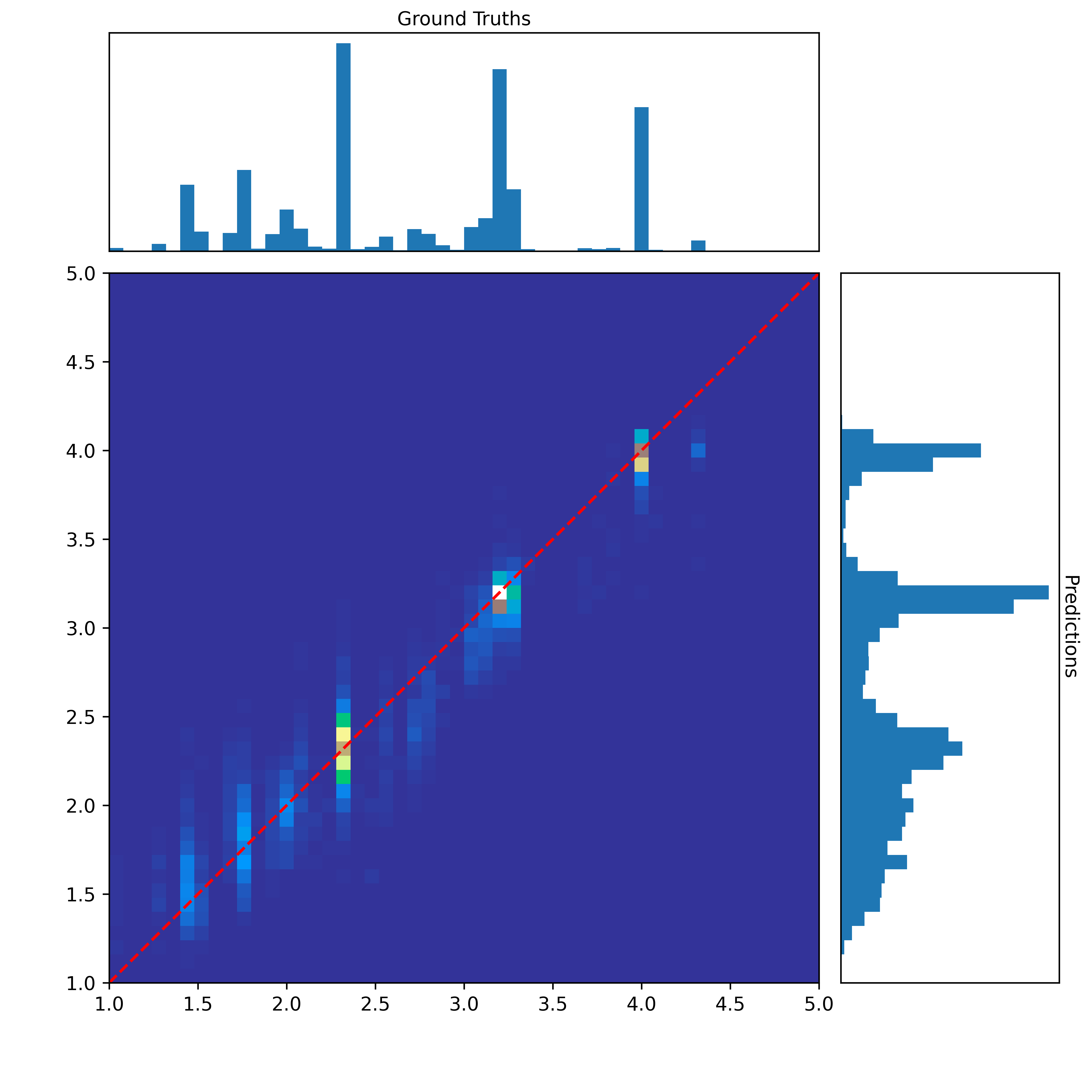}
        \caption{The ``proposed method w/ pretrain" model(\rts)}
        \label{fig:subfig32}
    \end{subfigure}
    \caption{Confusion Matrices for the best-performing models trained on \emph{Dataset II} in the ``Joint estimation of room parameters" task}
    \label{fig:volrt60}
\end{figure}

Fig. \ref{fig:volrt60} shows confusion matrices for volume and \rts\ parameter estimation in the ``Joint estimation of room parameters" task, highlighting the best-performing models. The x-axis and y-axis represent the log-10 exponent of room parameters (volume and \rts). From the visualization results, it can be observed that estimation performances of the CNN-based model, CRNN-based model, and the ``proposed method w/ pretrain" model gradually improves, and their fitting capabilities increase progressively.

The comprehensive analysis of experimental results in this study demonstrates the effectiveness of the joint estimation model for the blind room parameter estimation task. This method involves utilizing a single model to simultaneously estimate both room volume and \rts\ parameters, providing a more holistic understanding of acoustic environmental characteristics. Particularly, the ``proposed method w/ pretrain" model achieves the highest $\rho$ for both room volume and \rts\ parameters. This highlights the model's capability of capturing the intricate characteristics of acoustic environments through the joint estimation of room parameters.

\section{Conclusion and future work}

In this study, we aim to explore the feasibility of using attention-based models to address audio processing tasks, specifically including the ``Estimation of room volume parameter," ``Room parameter estimation under variable-length audio input," and ``Joint estimation of room parameters" tasks. We employ different training strategies to evaluate performances of a CNN-based model, a CRNN-based model, the proposed attention-based model, and the ``proposed method w/ pretrain" model.

Experimental results based on unseen real-world rooms and realistic noise scenario indicate that our proposed method shows significant superiority in terms of accurately capturing the acoustic characteristics of audio data. This demonstrates that neural networks based on pure attention mechanisms can effectively handle regression problems related to audio and exhibit potential advantages in handling joint estimation tasks and variable-length inputs.

Future research directions will focus on optimizing and enhancing the performance of attention-based audio processing models in real-world applications. We plan to further improve the model structure, including considering more efficient variants, to better capture the complex features of audio data. Additionally, we will strive to collect more comprehensive and diverse room data to enhance the model's generalization capabilities. We also aim to update robust and state-of-the-art \rts\ estimators \cite{antsalo2001estimation,jasa2009efficient,gotz2022neural} to obtain more accurate ground truth. These efforts will contribute to advancing the application of attention-based audio processing models in real-world scenarios.

\bibliography{sn-bibliography}

\end{document}